\documentclass[aps,prd,preprintnumbers,twocolumn,groupedaddress,nofootinbib,showpacs]{revtex4-1}
\usepackage{graphicx}
\usepackage{amsmath}
\usepackage{amssymb}
\usepackage{mathtools}
\usepackage{subfigure}
\usepackage{float} 
\usepackage{color}

\def\lsim{\mathrel{\raise.3ex\hbox{$<$\kern-.75em\lower1ex\hbox{$\sim$}}}}
\def\gsim{\mathrel{\raise.3ex\hbox{$>$\kern-.75em\lower1ex\hbox{$\sim$}}}}

\begin{document}

\title{Examining The Fermi-LAT Third Source Catalog In Search Of Dark Matter Subhalos}  
\author{Bridget Bertoni$^{1}$}
\author{Dan Hooper$^{2,3}$}
\author{Tim Linden$^4$}
\affiliation{$^1$Department of Physics, University of Washington, Seattle, WA 98195}
\affiliation{$^2$Center for Particle Astrophysics, Fermi National Accelerator Laboratory, Batavia, IL 60510}
\affiliation{$^3$Department of Astronomy and Astrophysics, University of Chicago, Chicago, IL 60637}
\affiliation{$^4$Kavli Institute for Cosmological Physics, University of Chicago, Chicago, IL 60637}

\date{\today}

\begin{abstract}

Dark matter annihilations taking place in nearby subhalos could appear as gamma-ray sources without detectable counterparts at other wavelengths. In this study, we consider the collection of unassociated gamma-ray sources reported by the Fermi Collaboration in an effort to identify the most promising dark matter subhalo candidates. While we identify 24 bright, high-latitude, non-variable sources with spectra that are consistent with being generated by the annihilations of $\sim$\,20-70 GeV dark matter particles (assuming annihilations to $b\bar{b}$), it is not possible at this time to distinguish these sources from radio-faint gamma-ray pulsars. Deeper multi-wavelength observations will be essential to clarify the nature of these sources. It is notable that we do not find any such sources that are well fit by dark matter particles heavier than $\sim$100 GeV. We also study the angular distribution of the gamma-rays from this set of subhalo candidates, and find that the source 3FGL J2212.5+0703 prefers a spatially extended profile (of width $\sim$\,$0.15^{\circ}$) over that of a point source, with a significance of 4.2$\sigma$ (3.6$\sigma$ after trials factor). Although not yet definitive, this bright and high-latitude gamma-ray source is well fit as a nearby subhalo of $m_{\chi} \simeq$\,\,20-50 GeV dark matter particles (annihilating to $b\bar{b}$) and merits further multi-wavelength investigation. Based on the subhalo distribution predicted by numerical simulations, we derive constraints on the dark matter annihilation cross section that are competitive to those resulting from gamma-ray observations of dwarf spheroidal galaxies, the Galactic Center, and the extragalactic gamma-ray background.

\end{abstract}

\pacs{95.35.+d, 95.95.Pw, 07.85.-m, FERMILAB-PUB-15-124-A}
\maketitle

\section{Introduction}

Numerical simulations reveal that dark matter structures form hierarchically, beginning with the smallest halos, and gradually merging to create ever larger systems, including the halos that host galaxies and galaxy clusters~\cite{White:1991mr}.  As a consequence of this process, dark matter halos are predicted to contain very large numbers of smaller subhalos. In the case of the Milky Way, the largest members of this subhalo population include the few dozen known dwarf galaxies and the Large and Small Magellanic Clouds. These systems are exceptional, however, and reflect only the very small fraction of subhalos that were large enough to capture significant quantities of gas and form stars. The vast majority of the Milky Way's subhalos harbor no significant quantities of baryonic matter and cannot be detected by optical surveys. If dark matter particles annihilate at a rate similar to that expected of a simple thermal relic, however, the nearest and most massive subhalos could generate an observable flux of gamma-rays, appearing as a population of unidentified gamma-ray sources.

The Fermi Collaboration has recently released a new catalog of gamma-ray sources, known as the Fermi-LAT Third Source Catalog, or 3FGL~\cite{TheFermi-LAT:2015hja}. Along with many identified objects, this catalog contains 992 sources that have not been associated with emission observed at other wavelengths. These sources almost certainly include many presently unidentified blazers, pulsars, and other astrophysical objects.\footnote{In the Galactic Plane, some 3FGL sources also remain unassociated due to challenges in discriminating between multiple astrophysical objects in the same region of the sky.} If the dark matter consists of annihilating particles with weak-scale masses, however, we should also expect a relatively small number of these unassociated sources to be dark matter subhalos~\cite{Kuhlen:2008aw,Pieri:2007ir}. For example, we estimate that for a 100 GeV dark matter particle with an annihilation cross section of $\sigma v \simeq 2\times 10^{-26}$ cm$^3$/s, the 3FGL should contain on the order of $\sim$10 sources which are, in fact, dark matter subhalos.  

The challenge, of course, lies not in merely detecting such subhalos, but in differentiating them from the much more numerous conventional unidentified sources. With this goal in mind, we expand upon previous work~\cite{Berlin:2013dva,Belikov:2011pu,Buckley:2010vg,Zechlin:2011wa,Mirabal:2012em,Mirabal:2010ny,Zechlin:2012by} in a number of ways. First of all, we employ Fermi's recently recently catalog of gamma-ray sources, the 3FGL~\cite{TheFermi-LAT:2015hja}. We also make use of the current Fermi dataset to extract a finely binned spectrum from each unassociated source, allowing us to make detailed comparisons with the predictions of various dark matter models. From this information, we place constraints on the dark matter annihilation cross section that are competitive with those derived from dwarf galaxies~\cite{Drlica-Wagner:2015xua,Geringer-Sameth:2014qqa}, the Galactic Center~\cite{Hooper:2012sr}, and the extragalactic gamma-ray background~\cite{Ackermann:2015tah,DiMauro:2015tfa}. We also identify a collection of 24 bright and high-latitude gamma-ray sources with dark matter-like spectra.  We further investigate this list of prospective dark matter subhalo candidates by testing their angular distribution, and find evidence of spatial extension from the subhalo candidate source 3FGL J2212.5+0703.

The remainder of this paper is structured as follows. In Sec.~\ref{review}, we describe our calculation of the distribution of dark matter subhalos and their predicted gamma-ray fluxes. In Sec.~\ref{3fgl}, we discuss the characteristics of the unassociated source population presented in the 3FGL catalog, and place cuts on variability and galactic latitude in an effort to separate prospective subhalo candidates from astrophysical sources. In Sec.~\ref{data}, we describe our analysis of the Fermi data and the determination of the spectra from the 3FGL's unassociated sources. Focusing on the most promising subhalo candidates, we test these sources for indications of spatial extension in Sec.~\ref{extension}. In Sec.~\ref{constraints}, we use the number of observed subhalo candidates to derive constraints on the dark matter annihilation cross section, finding limits that are competitive with the strongest constraints from other gamma-ray observations. Finally, in Sec.\ref{summary}, we summarize our results and comment on the prospects for future study.

\section{Gamma-Rays From Nearby Dark Matter Subhalos}
\label{review}

The prospects for observing gamma-rays from nearby dark matter subhalos depend not only on the characteristics of the dark matter particle itself, but on the local abundance and density profiles of the subhalos. To estimate these quantities, we make use of the results of the Aquarius Project, which has provided the highest resolution simulations to date of the dark matter subhalo populations found within the halos of Milky Way-like galaxies, identifying hundreds of thousands of subhalos, with masses as low as $3.24\times 10^4 M_{\odot}$~\cite{Springel:2008cc}. The mass distribution of these subhalos takes the form of $dN/dM \propto M^{-1.9}$, with an overall normalization that corresponds to 13.2\% of the Milky Way's total mass in dark matter. 
 
In our calculations, we follow the approach of Ref.~\cite{Berlin:2013dva}, to which we direct the reader for further details. We will, however, repeat the main elements of our calculation here. Firstly, based on the results of the Aquarius simulation, we adopt the following distribution for subhalos in the local volume of the Milky Way:
\begin{equation}
\frac{dN}{dM dV} = 260 \,\, {\rm kpc}^{-3} \, M_{\odot}^{-1} \times \bigg(\frac{M}{M_{\odot}}\bigg)^{-1.9}.  
\label{norm}
\end{equation}
By integrating this expression between $M = 3.24 \times 10^4 \, M_{\odot}$ and $10^7 M_{\odot}$, we find a local mass density in subhalos of 5700 $M_{\odot}/$kpc$^3$ (0.00022 GeV/cm$^3$), corresponding to approximately 0.054\% of the overall local dark matter density (in good agreement with Fig.~12 of Ref.~\cite{Springel:2008cc}).

For each individual subhalo, we adopt an Einasto profile with $\alpha=0.16$, truncated by the effects of tidal stripping, such that only the innermost 0.5\% of the mass remains intact~\cite{Berlin:2013dva,Springel:2008cc}. For the initial concentration of each subhalo (prior to tidal effects), defined as the ratio of the virial and scale radii, $c \equiv r_{\rm vir}/r_{-2}$, we adopt values as presented in Ref.~\cite{Sanchez-Conde:2013yxa}, with subhalo-to-subhalo variations modeled by a log-normal distribution with a dispersion of $\sigma_c=0.24$~\cite{Bullock:1999he}. In this respect, we depart from the mass-concentration relationship used in Ref.~\cite{Berlin:2013dva}, as based on the results of Ref.~\cite{MunozCuartas:2010ig}. This update of the mass-concentration relationship reduces the number of detectable subhalos (in gamma-rays) by a factor of approximately 4.6 relative to that presented in Ref.~\cite{Berlin:2013dva}.

The population of subhalos detectable in gamma-rays is dominated by the most massive and nearby members of this population. In our calculations, we include subhalos with masses up to $10^7 M_{\odot}$. We have chosen to limit our calculation to subhalos below this mass because we expect many of the more massive subhalos to contain significant quantities of baryons (stars and/or gas) and thus would be identified with dwarf spheroidal galaxies.

\begin{figure}
\vspace{-3.0cm}
\mbox{\includegraphics[width=0.52\textwidth]{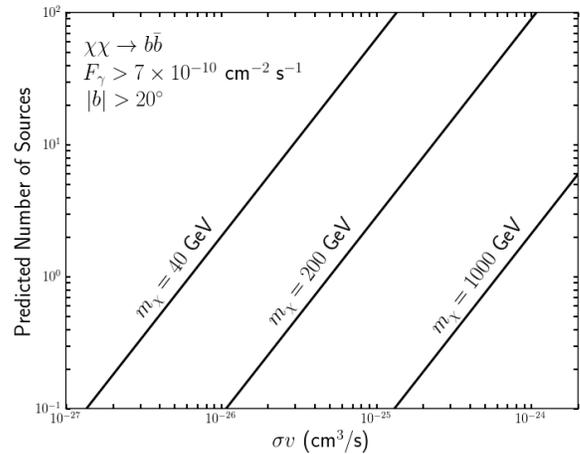}} 
\vspace{-3.0cm}
\caption{The number of high latitude ($|b|>20^{\circ}$) dark matter subhalos predicted by our calculations to be bright gamma-ray sources ($F_{\gamma} > 7 \times 10^{-10}$ cm$^{-2}$ s$^{-1}$, $E_{\gamma} > 1$ GeV), as a function of the annihilation cross section, for three values of the mass. Here, we have assumed annihilations that proceed to $b\bar{b}$.}
\label{detect}
\end{figure}

\begin{figure*}
\vspace{-3.0cm}
\mbox{\includegraphics[width=0.52\textwidth]{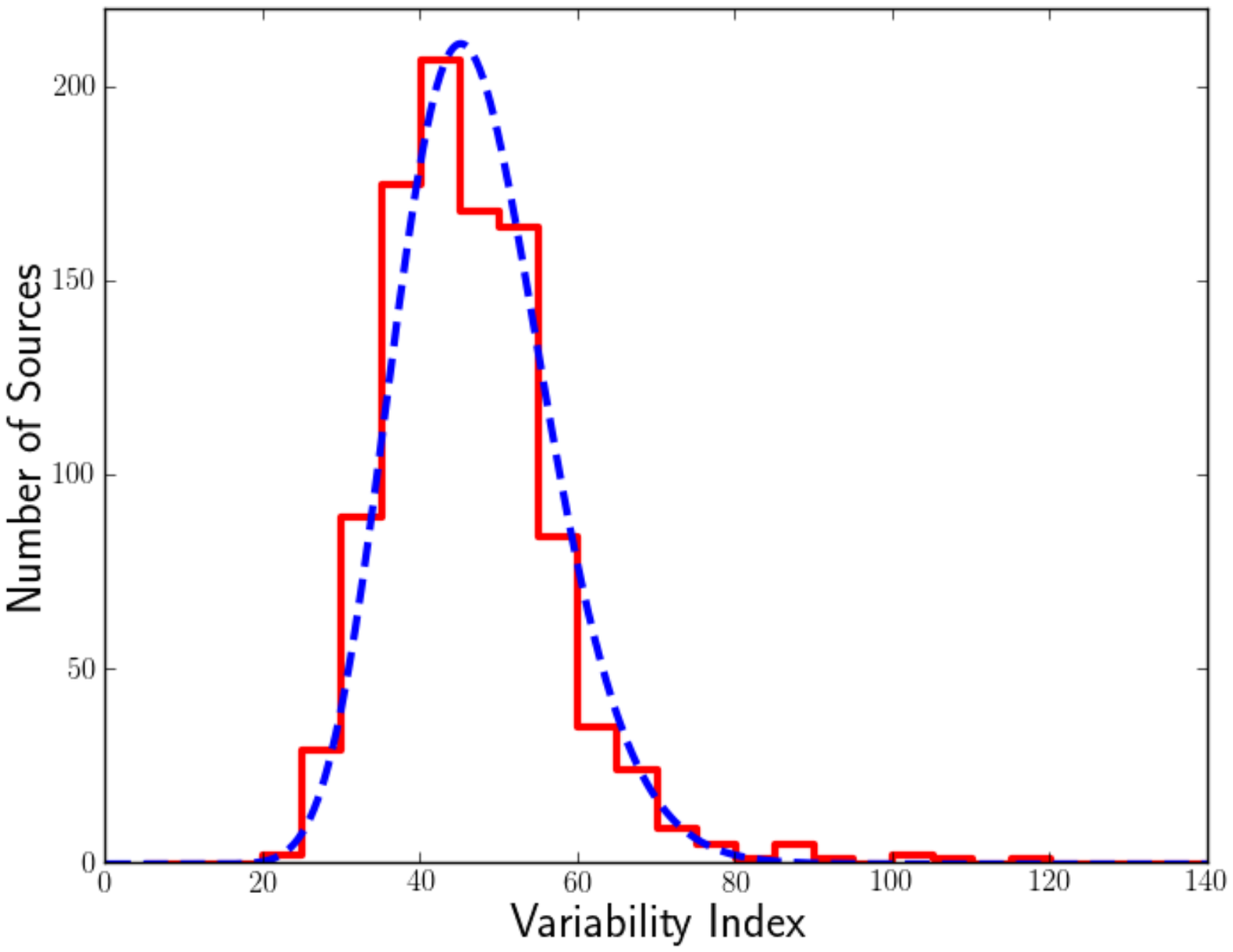}}
\hspace{-1.0cm}
\mbox{\includegraphics[width=0.52\textwidth]{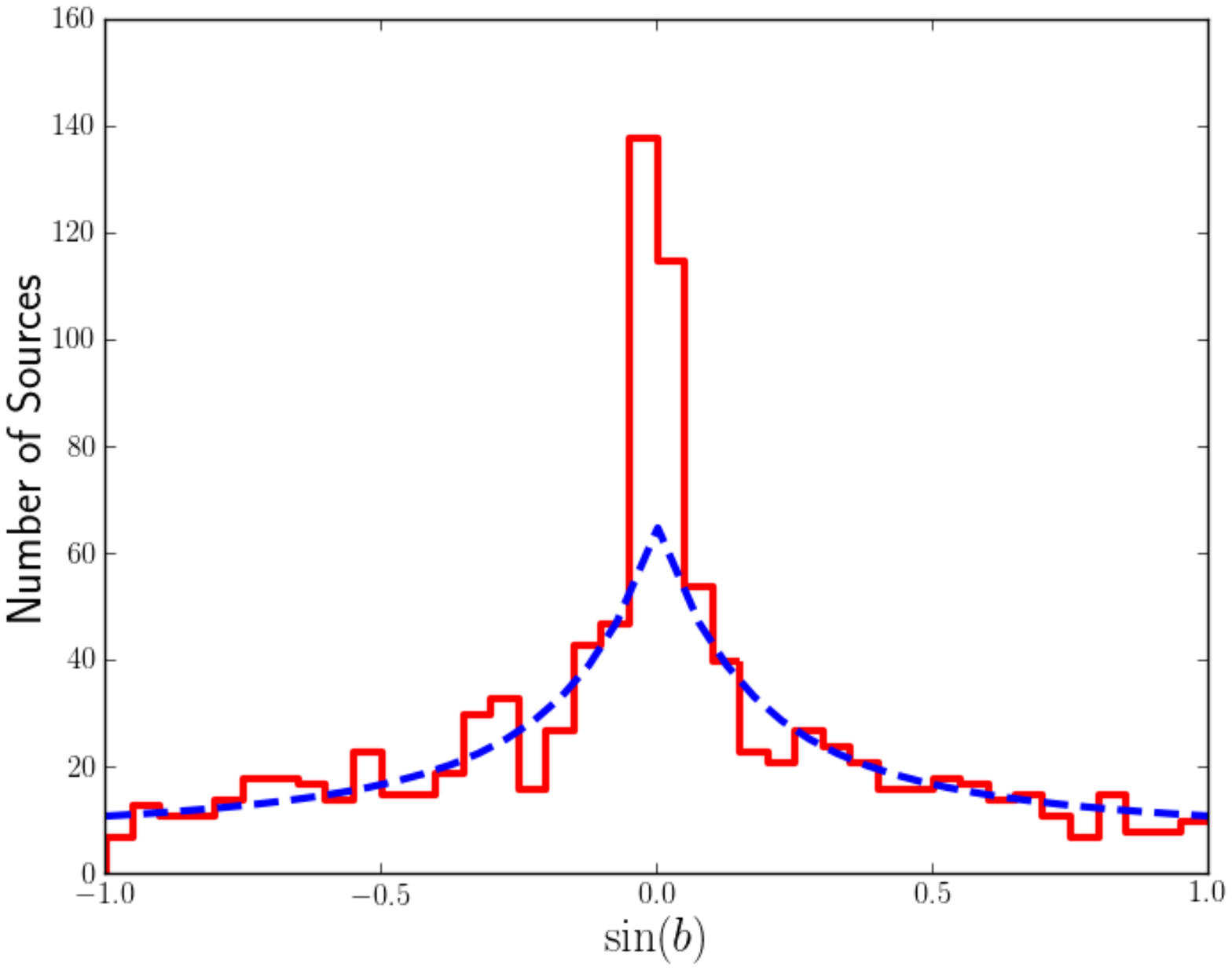}}
\vspace{-3.2cm}
\caption{Left: The distribution of the variability index observed among the 3FGL's unassociated source population. The dashed curve represents the statistical distribution predicted for a population of non-variable sources (a $\chi^2$ distribution with 47 degrees-of-freedom). Right: The latitude distribution of the 3FGL's unassociated sources. Based on a simple disk+isotropic population model (dashed curve), we estimate that approximately 20\% of these sources are part of an isotropic population. In our analysis, we will focus on the unassociated sources with a variability index less than 80 and that are located at $|b|>20^{\circ}$ ($|\sin b | > 0.342$).}
\label{varlat}
\end{figure*}

The differential gamma-ray spectrum per solid angle from dark matter annihilations within an individual subhalo is given by:
\begin{align}
\label{flux1}
\Phi (E_{\gamma}, \theta) = \frac{1}{8 \pi  m_{\chi}^2}  \langle \sigma v \rangle \frac{\mathrm{d}N_{\gamma}}{\mathrm{d}E_\gamma}  \,  \int_{\text{l.o.s.}} \rho^2 [ r(D,l,\theta)] \,\mathrm{d}l, 
\end{align}
where $m_{\chi}$ is the mass of the dark matter particle, $\langle \sigma v \rangle$ is the annihilation cross section, and $dN_{\gamma}/dE_{\gamma}$ is the gamma-ray spectrum produced per annihilation, which we calculate using PYTHIA 8~\cite{pythia}. The integral of the density squared is performed over the line-of-sight, $D$ is the distance to the center of the subhalo, $\theta$ is the angle to the center of the subhalo, and $r(\theta,D,l)=\sqrt{D^2+l^2-2Dl\cos{\theta}}$.

In Fig.~\ref{detect}, we plot the number of high latitude ($|b|>20^{\circ}$) dark matter subhalos that are predicted to be bright gamma-ray sources ($F_{\gamma} > 7 \times 10^{-10}$ cm$^{-2}$ s$^{-1}$, $E_{\gamma} > 1$ GeV, which is well above the threshold for inclusion in the 3FGL catalog), as a function of the annihilation cross section, for three choices of the mass (for the case of annihilations proceeding to $b\bar{b}$). For an annihilation cross section near the value predicted for a simple thermal relic ($\sigma v \sim 2\times 10^{-26}$ cm$^3$/s), one expects such bright subhalos to exist only for relatively low values of the dark matter mass, $m_{\chi} \lsim 100$ GeV. For lighter masses, however, one predicts that Fermi should have already detected gamma-ray emission from several such subhalos, constituting a population of presently unidentified sources without associated emission at other wavelengths.

We emphasize that although non-negligible uncertainties do exist in our calculation of the number of dark matter subhalos above a given gamma-ray flux, the overall conclusions of these calculations are generally robust. Reasonable changes in our assumptions regarding the local number density of subhalos and dark matter distribution within subhalos could plausibly increase or decrease the predicted number of gamma-ray sources by up to a factor of a few. That being said, we have made a number of choices which we consider to be conservative, and thus expect our calculations to reflect a fairly realistic, albeit somewhat low, estimate for the number of such sources that might be observed. For example, steeper density profiles (as suggested by the Via Lactea II simulation~\cite{Diemand:2009bm}) could increase the predicted number of observable gamma-ray subhalos. Furthermore, if we had included subhalos larger than our cutoff of $10^7 M_{\odot}$, greater numbers of such sources would be predicted. We have also neglected any annihilation ``boost factors'' that might result from substructure within individual dark matter subhalos.

\section{Unidentified Sources In the 3FGL Catalog}
\label{3fgl}

The Fermi Collaboration has recently released their third source catalog (3FGL), based on their first four years of data~\cite{TheFermi-LAT:2015hja}. This catalog contains 3033 sources, each detected with greater than approximately 4$\sigma$ significance. About half of these sources have been identified as, or associated with, known active galaxies (including BL Lacs, flat spectrum radio quasars, and other classes of active galaxies). A smaller, but not insignificant number, of these sources have also been associated with galactic objects, including pulsars, supernova remnants, and globular clusters.  Of the 3033 sources contained in the 3FGL, 992 have not yet been identified or associated with emission observed at other wavelengths. It is among this subset of 3FGL sources that we could potentially find a population of dark matter subhalos. 

The 3FGL provides information about each source that we will use to refine our search for dark matter subhalos. First, the Fermi Collaboration has tested each source for variability, as can be exhibited by some classes of astrophysical sources, but not by dark matter subhalos.  Note that this test is performed by dividing the data into month-long temporal bins, and thus is not sensitive to variations taking place over shorter timescales, such as is observed from pulsars, for example. For each source, the Fermi Collaboration reports a value for its ``variability index''. In the left frame of Fig.~\ref{varlat}, we plot the distribution of this quantity observed among the 3FGL's unassociated source population. The dashed curve represents the statistical distribution predicted for a population of non-variable sources (a $\chi^2$ distribution with 47 degrees-of-freedom)~\cite{TheFermi-LAT:2015hja}. For variability indices lower than approximately 80, these results are in good agreement, and thus provide no evidence of a variable population. In addition, we identify 18 unassociated 3FGL sources with a variability index greater than 80. In our analysis, we remove these 18 sources from our list of potential subhalo candidates.

In contrast to other galactic gamma-ray sources, which tend to be concentrated near the Galactic Plane, bright dark matter subhalos are predicted to be approximately isotropically distributed on the sky. We can therefore use galactic latitude as an indicator of the likelihood that a given gamma-ray source is a dark matter subhalo. In the right frame of Fig.~\ref{varlat}, we plot the latitude distribution of the 3FGL's unassociated sources. As a histogram binned in $\sin(b)$, an isotropic distribution would be flat, whereas the observed distribution clearly includes a population of sources that is concentrated around the Galactic Plane.

To estimate how this distribution breaks down into different spatial populations, we model the total distribution of sources as the sum of an isotropic component and a component with a thick disk-like distribution.  For the disk-like distribution, we adopt $n \propto \exp(-z/z_0)$, and assume simply that individual sources are detectable out to a common distance of $d_{\rm max}$. Allowing the normalizations of the isotropic and disk-like components to vary, along with the parameters $z_0$ and $d_{\rm max}$, we find that the observed distribution is best-fit by $z_0/d_{\rm max} \simeq 0.1$ and $n_{\rm isotropic}/n_{\rm total} \simeq 0.45$ at $b=90^{\circ}$ (shown as a dashed curve in the right frame of Fig.~\ref{varlat}). In this fit, we have neglected the innermost two bins of the distribution, which are highly biased by the incompleteness of multi-wavelength AGN catalogs in this region of the sky~\cite{TheFermi-LAT:2015hja}.  This simple model suggests that roughly $\sim$20\% or $\sim$200 of Fermi's 992 unassociated sources are part of an isotropic population, consisting of extragalactic sources, and perhaps a small number of nearby dark matter subhalos. In our analysis, we will focus on Fermi's unassociated sources located at $|b|>20^{\circ}$, allowing us to limit contamination from galactic astrophysical sources.

\section{Fermi Data Analysis}
\label{data}

\begin{figure*}
\vspace{-1.8cm}
\mbox{\includegraphics[width=0.49\textwidth]{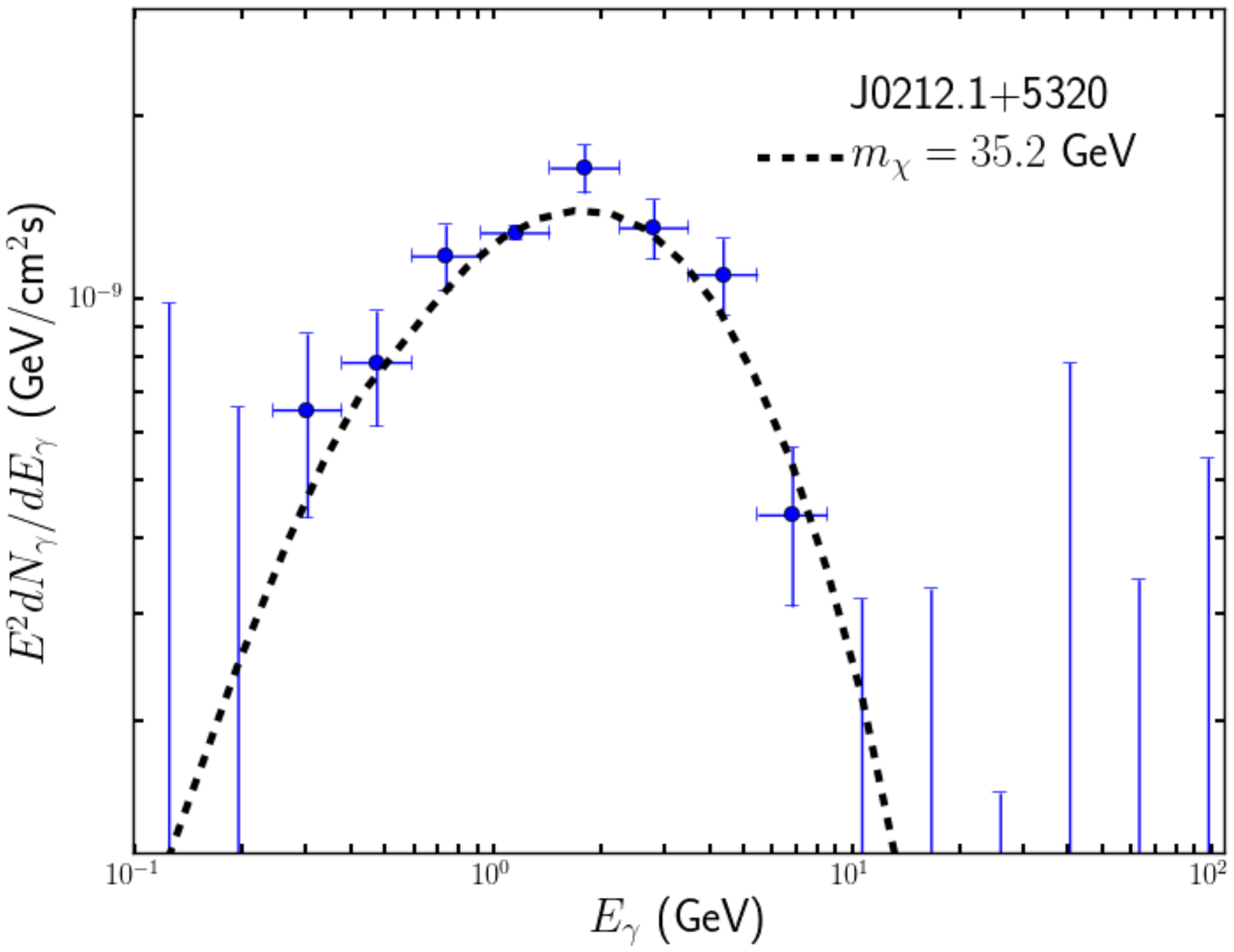}} 
\mbox{\includegraphics[width=0.49\textwidth]{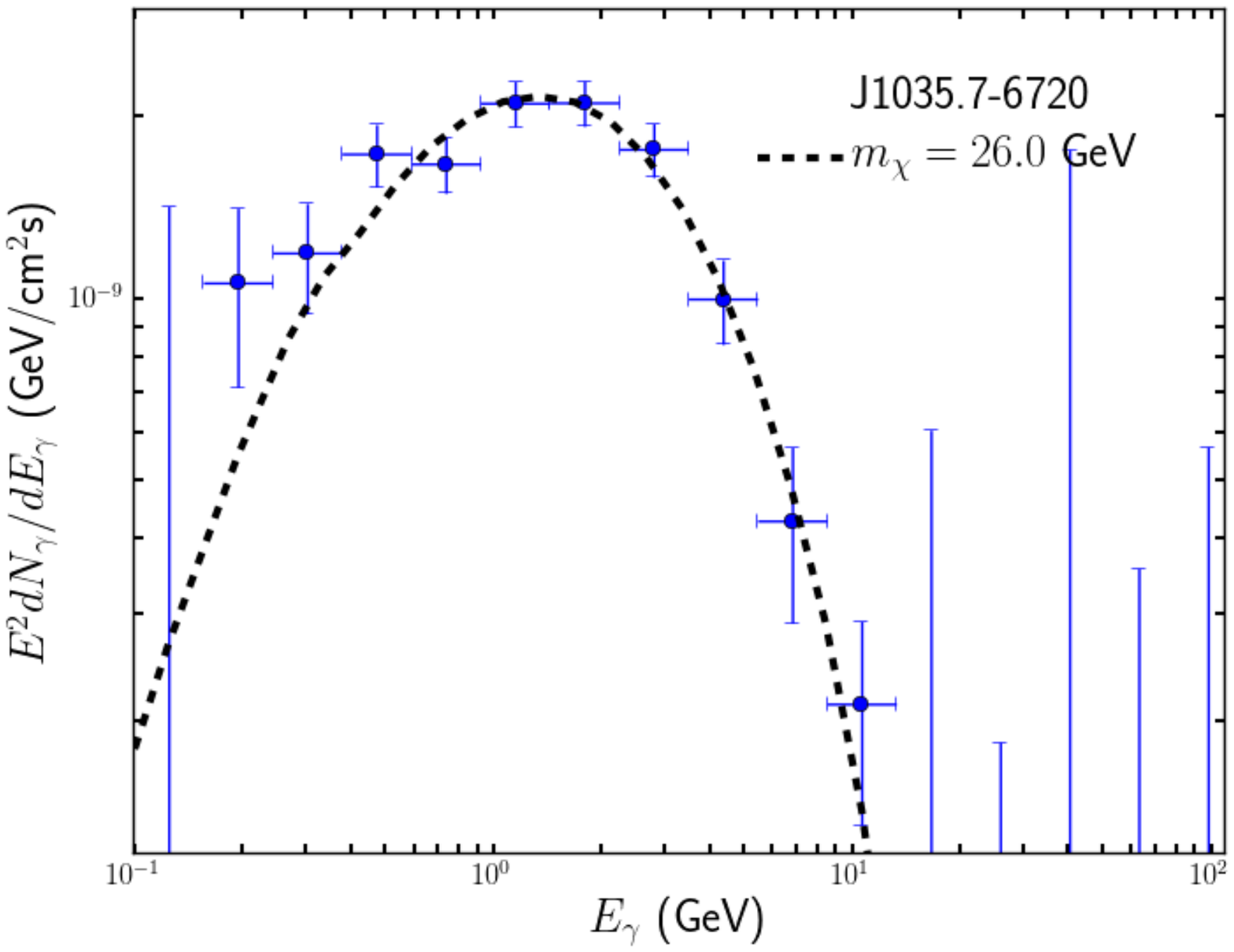}} \\
\vspace{-5.2cm}
\mbox{\includegraphics[width=0.49\textwidth]{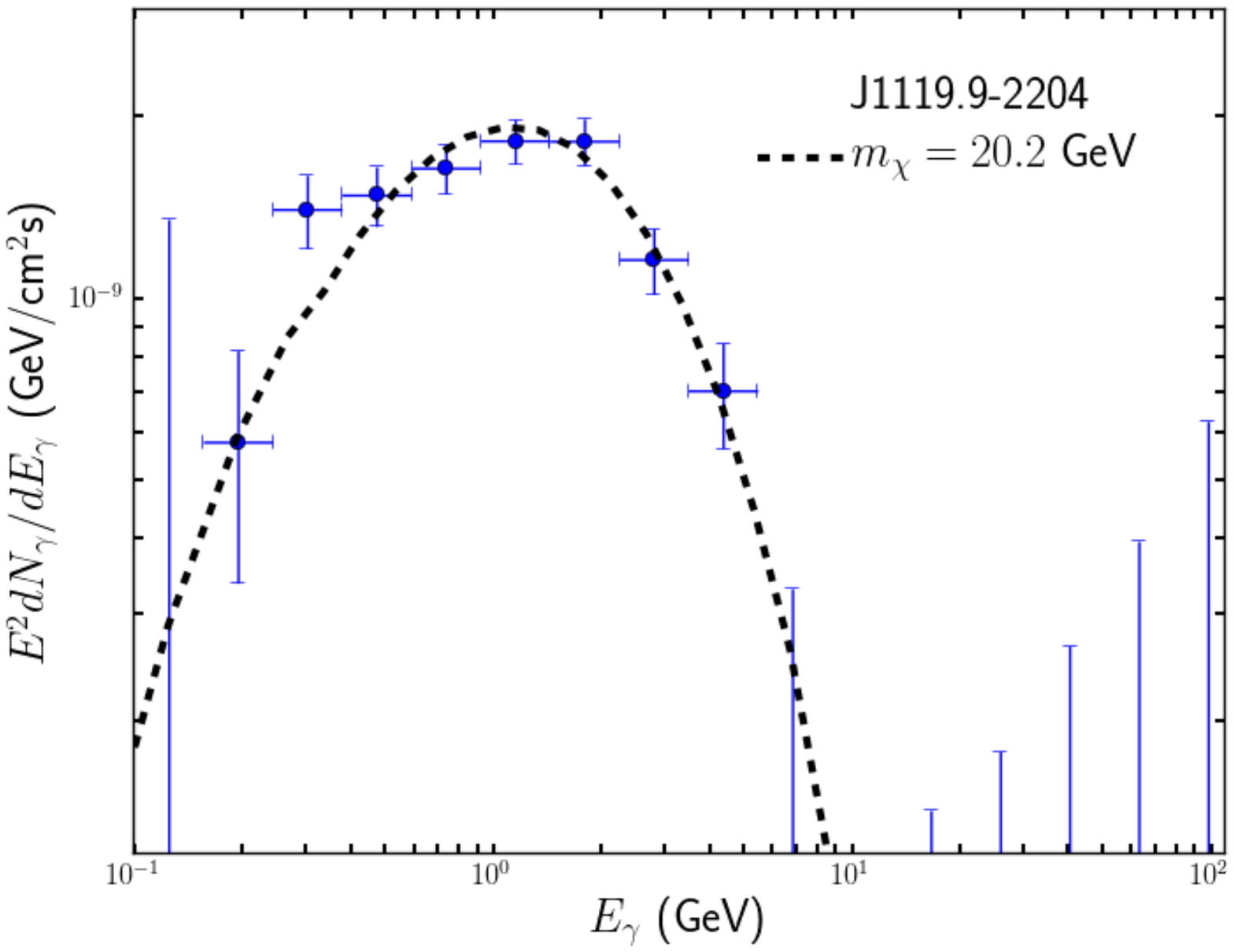}} 
\mbox{\includegraphics[width=0.49\textwidth]{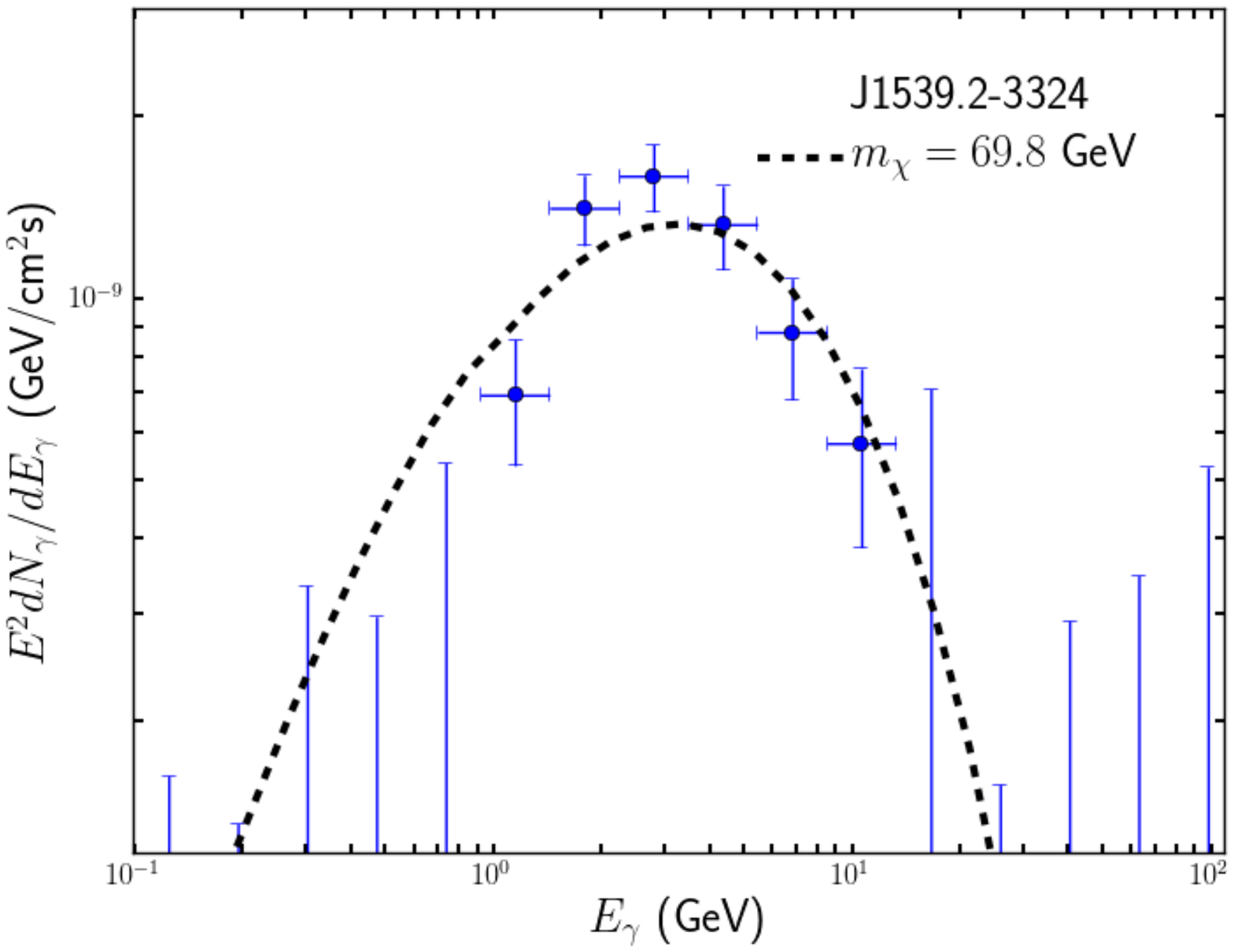}} \\ 
\vspace{-5.2cm}
\mbox{\includegraphics[width=0.49\textwidth]{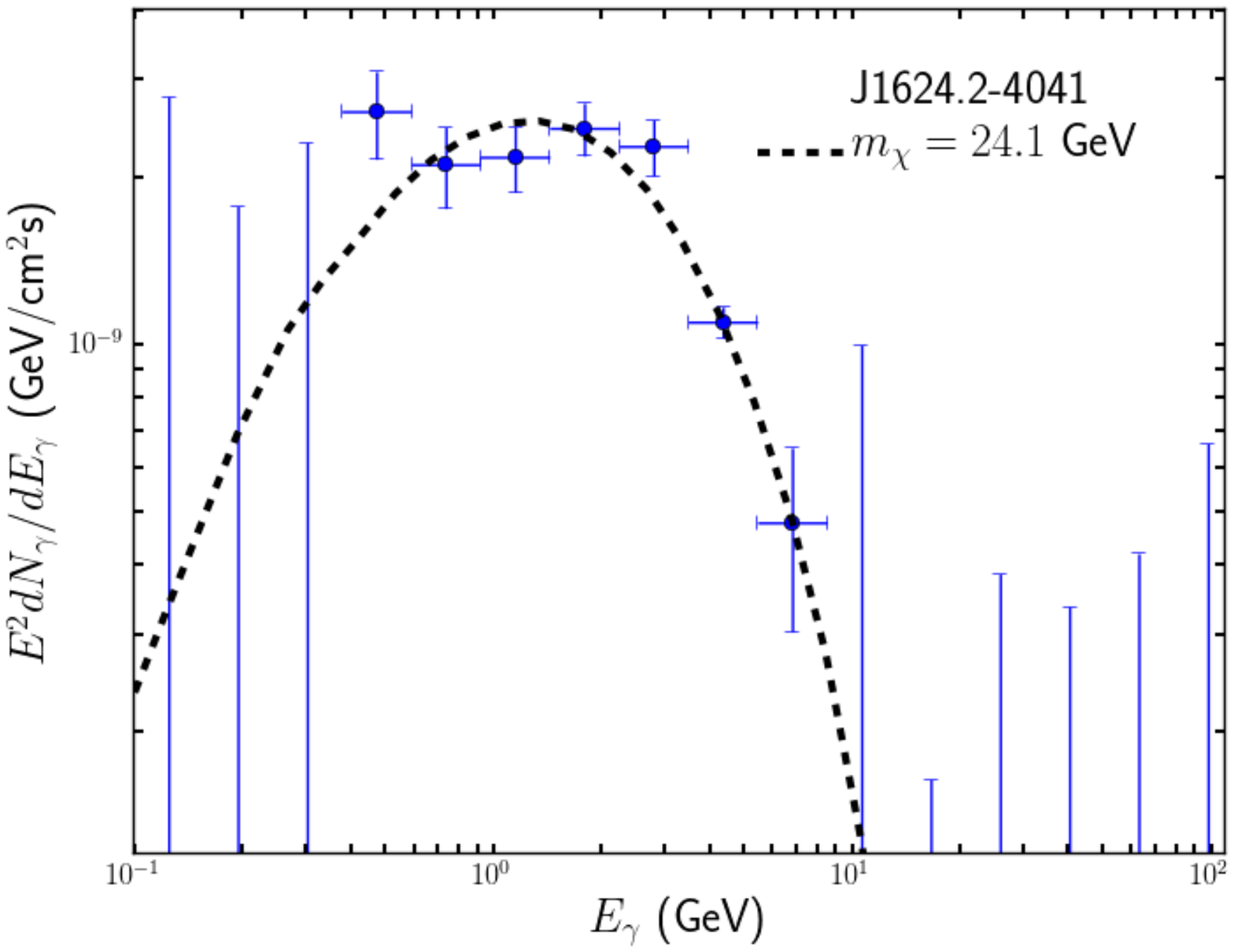}} 
\mbox{\includegraphics[width=0.49\textwidth]{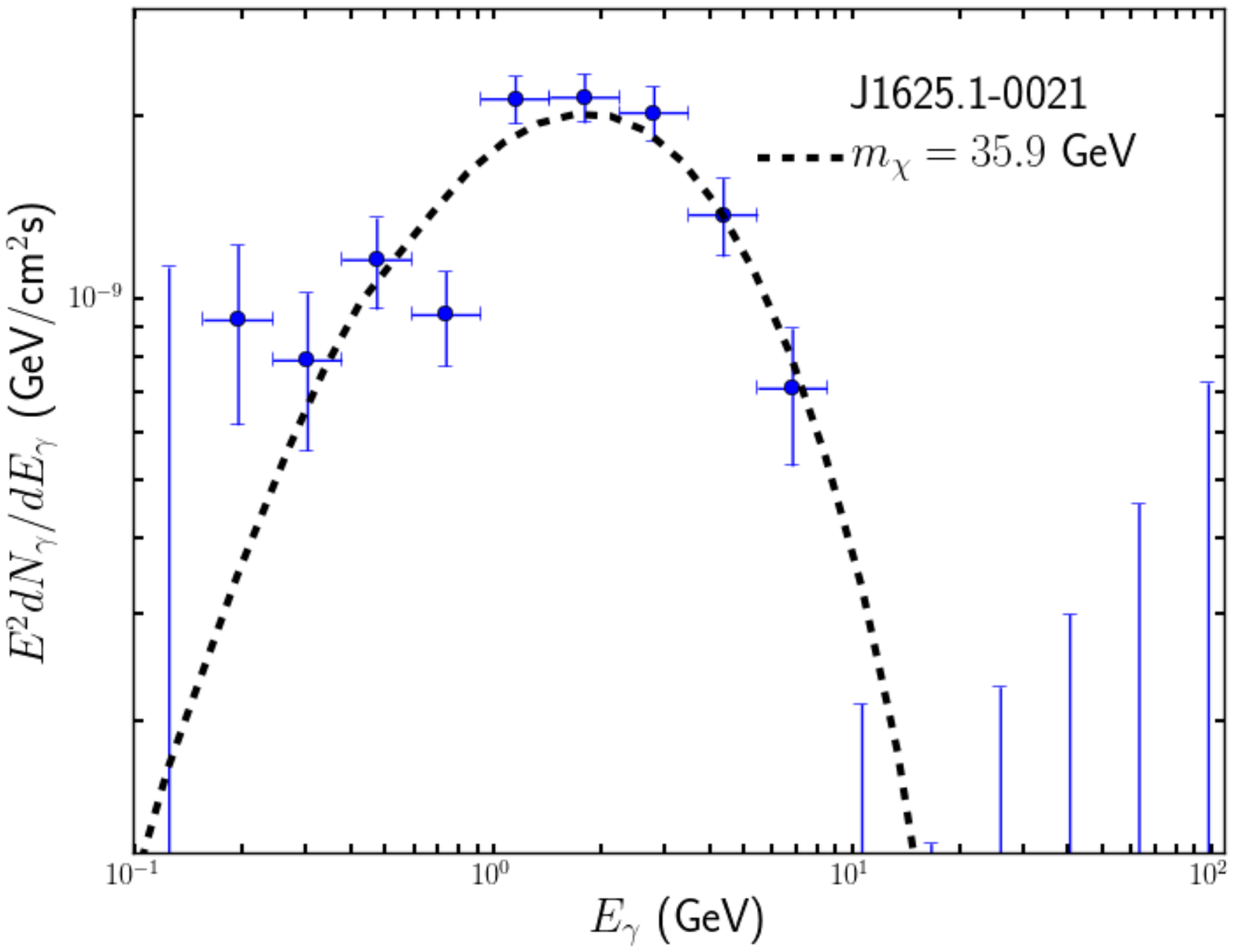}} \\
\vspace{-2.6cm}
\caption{The gamma-ray spectra from 9 of the brightest ($F_{\gamma} > 2 \times 10^{-9}$ cm$^{-2}$ s$^{-1}$, $E_{\gamma} > 1$ GeV) non-variable (variability index $< 80$), unassociated 3FGL sources located outside of the Galactic Plane ($|b|>5^{\circ}$). Each of these 9 sources are reasonably well fit ($\chi^2 < 25$) by dark matter annihilating to $b\bar{b}$, and the dashed curves represent this prediction for the best-fit value of the dark matter mass.}
\label{goodfits}
\end{figure*}

\renewcommand{\thefigure}{\arabic{figure} (Cont.)}
\addtocounter{figure}{-1}

\begin{figure*}
\vspace{5.5cm}
\mbox{\includegraphics[width=0.49\textwidth]{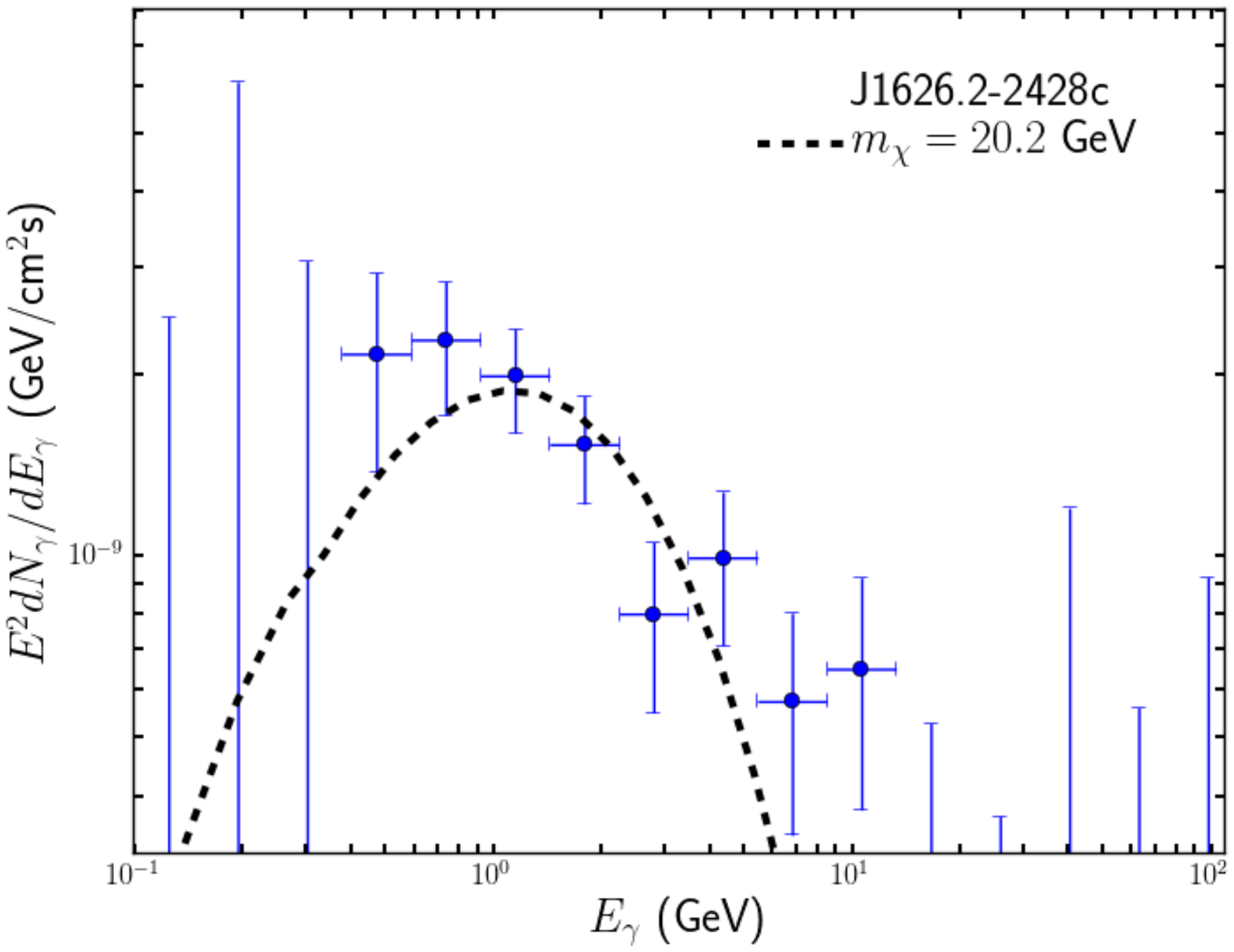}}  
\mbox{\includegraphics[width=0.49\textwidth]{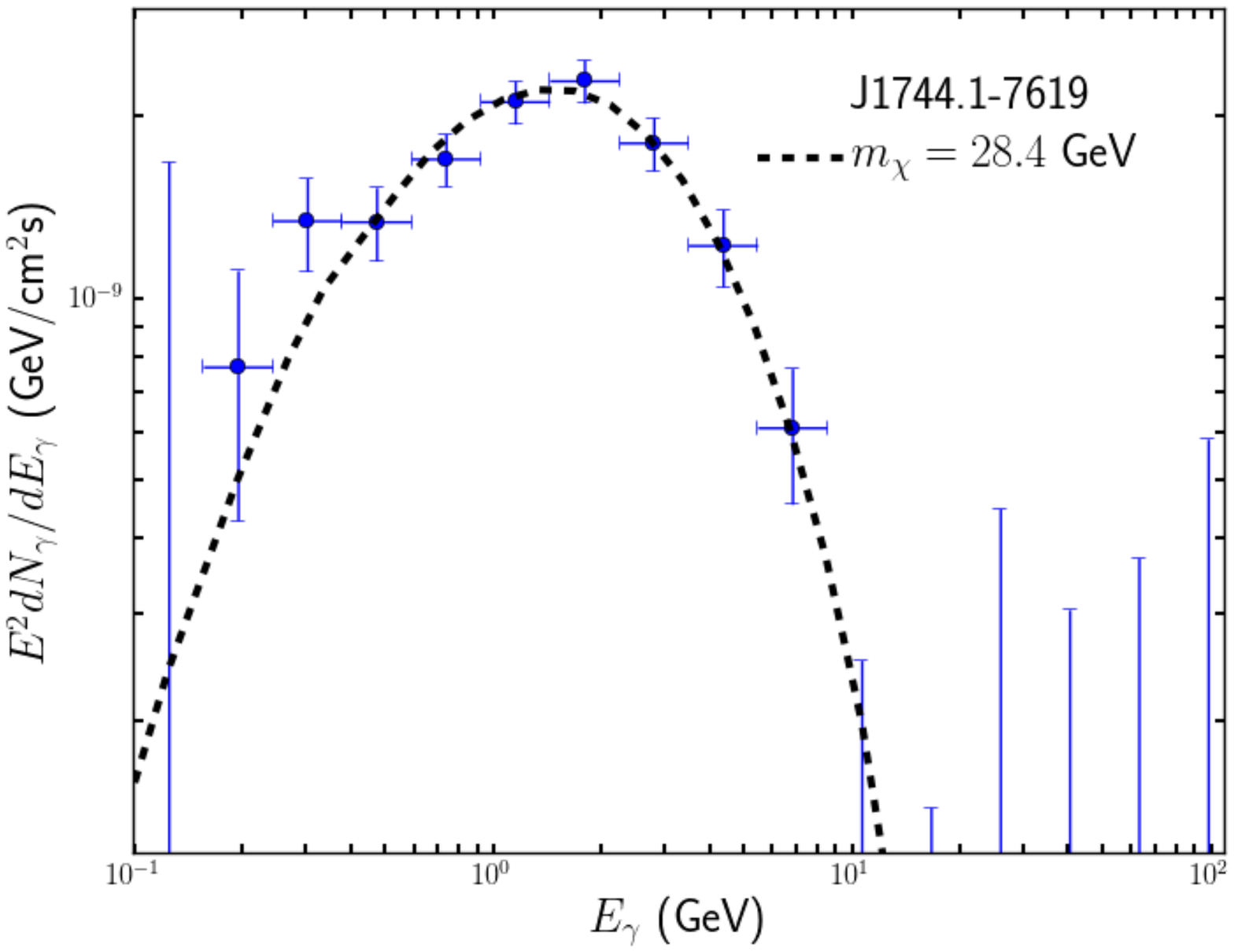}} \\
\vspace{-5.2cm}
\mbox{\includegraphics[width=0.49\textwidth]{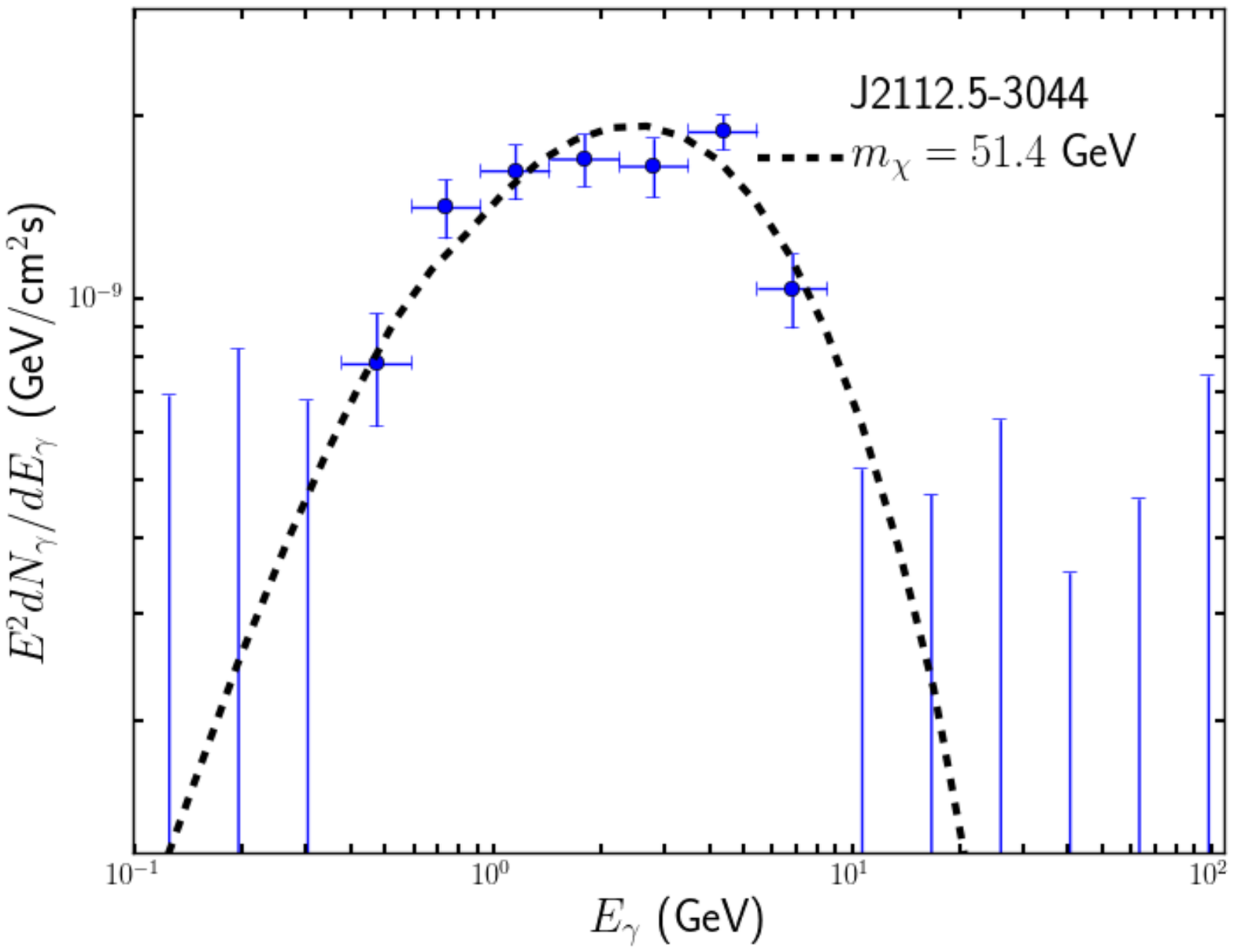}} \\
\vspace{-2.6cm}
\caption{}
\end{figure*}

\renewcommand{\thefigure}{\arabic{figure}}

\begin{figure*}
\vspace{-1.5cm}
\mbox{\includegraphics[width=0.49\textwidth]{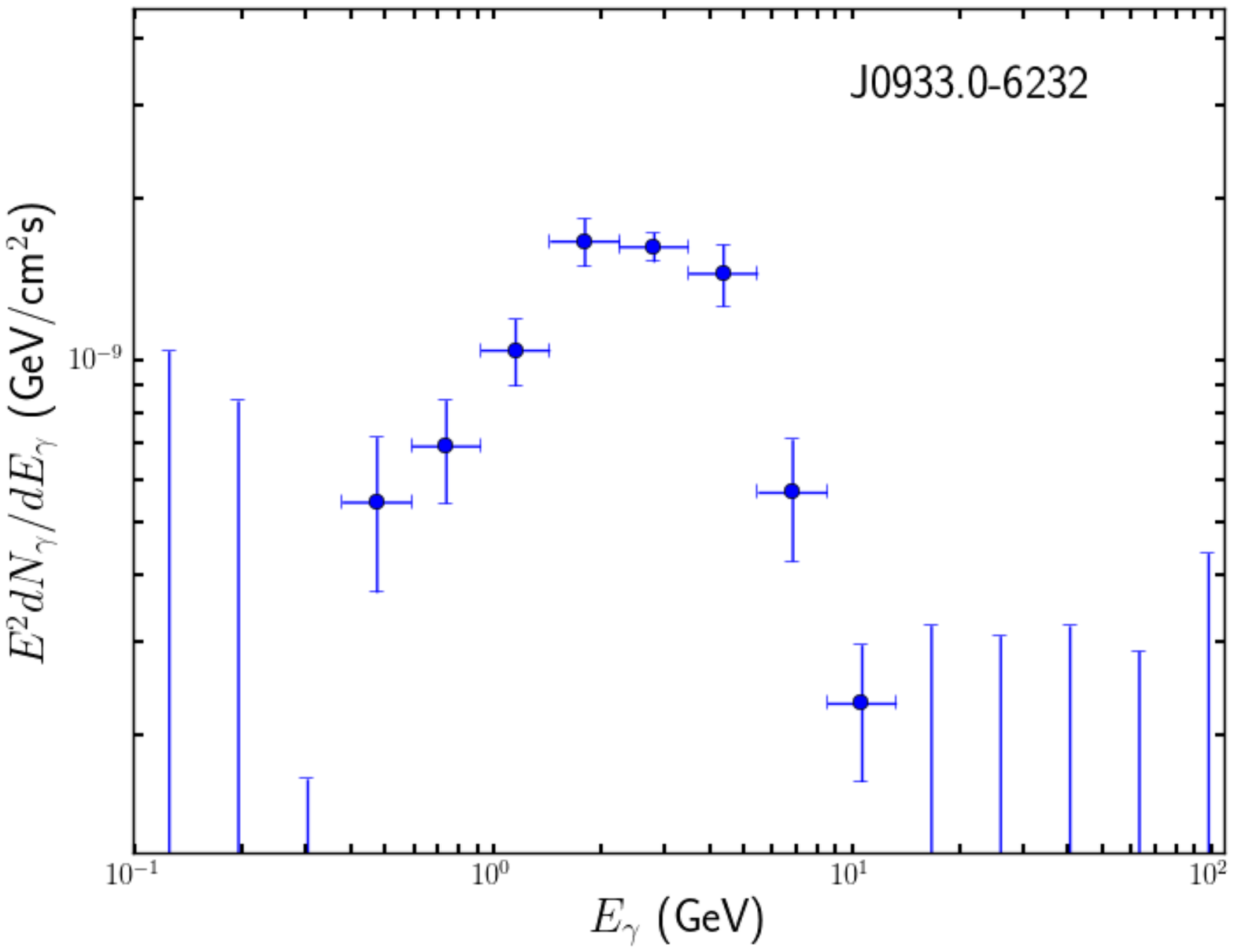}} 
\mbox{\includegraphics[width=0.49\textwidth]{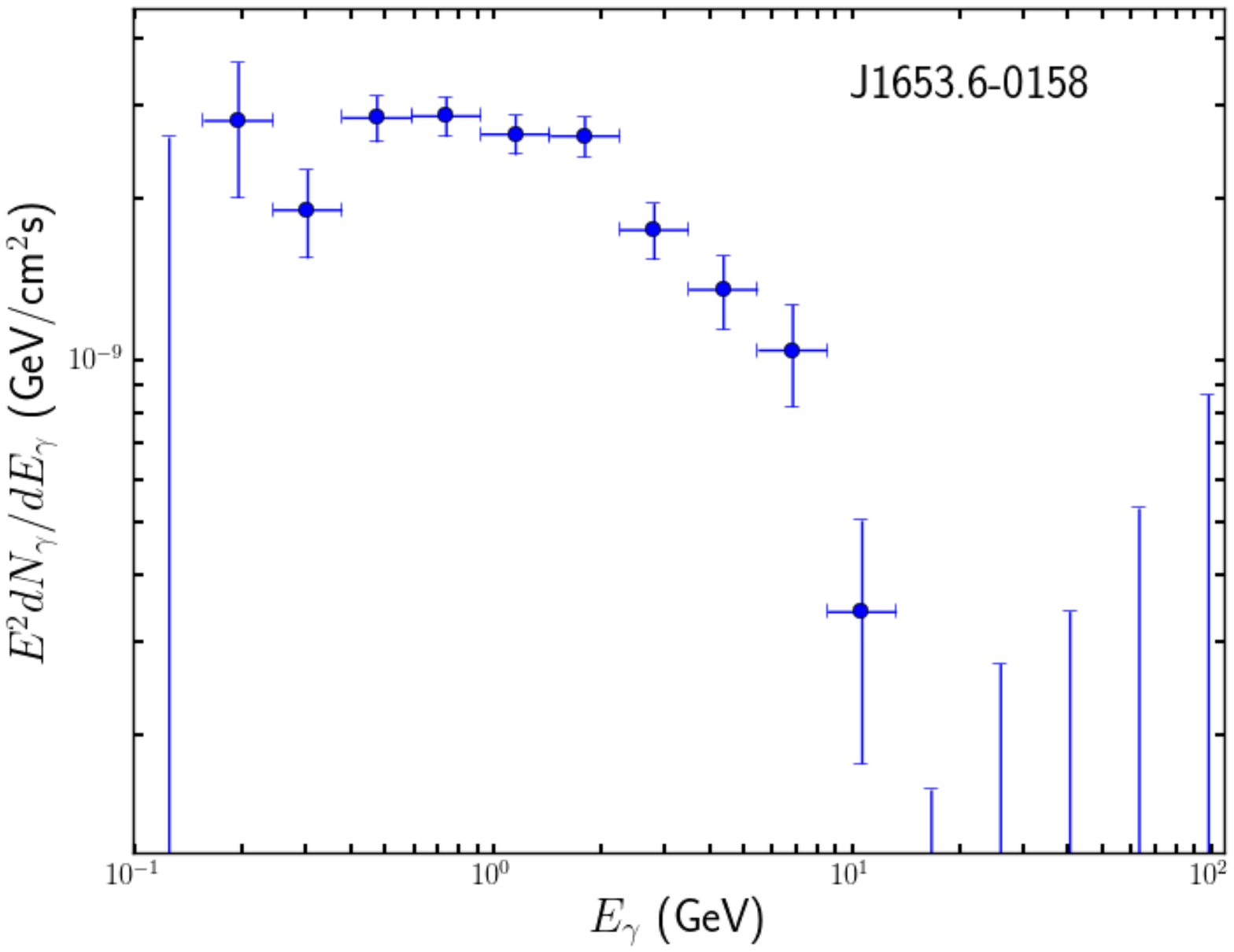}} \\
\vspace{-5.2cm}
\mbox{\includegraphics[width=0.49\textwidth]{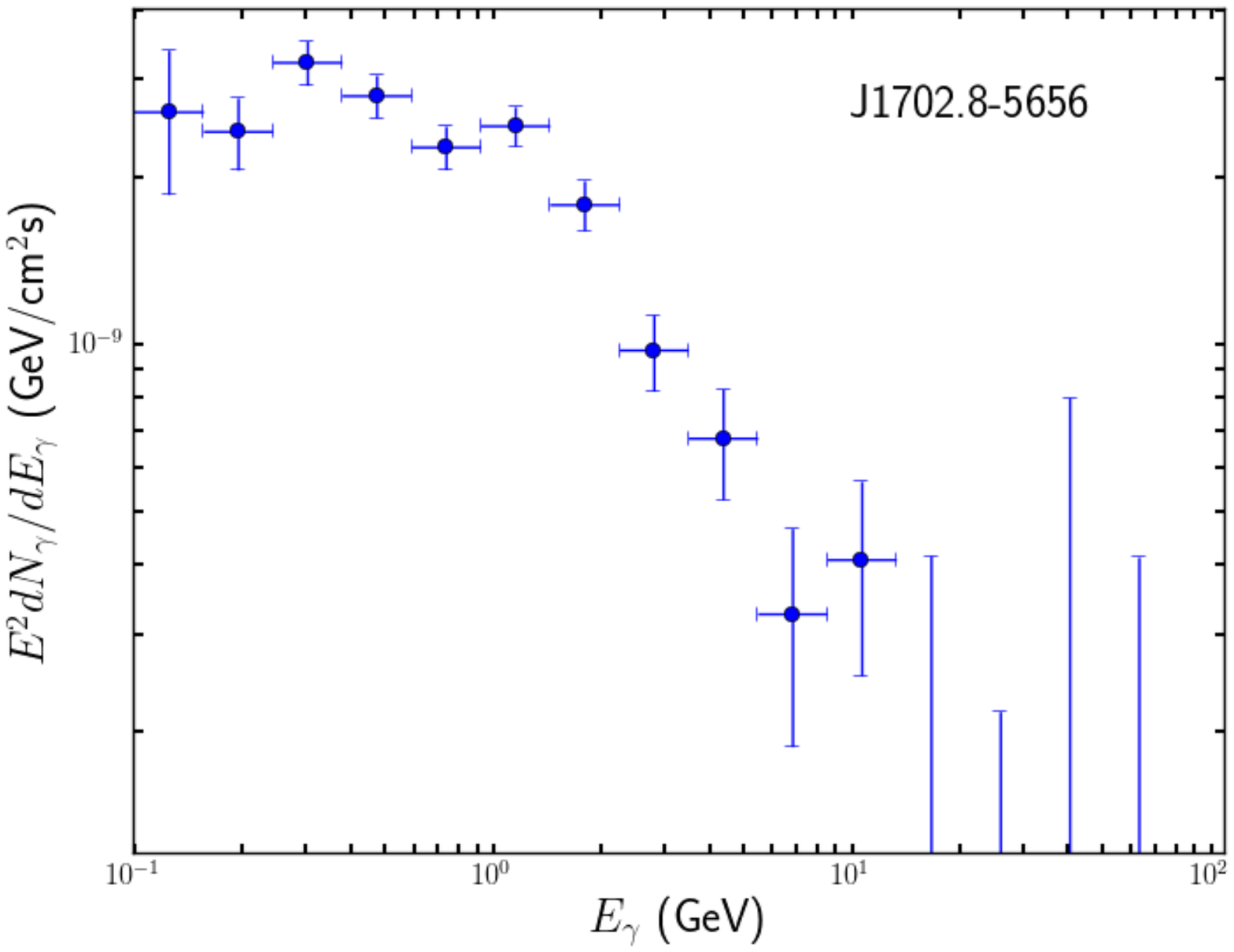}}  
\mbox{\includegraphics[width=0.49\textwidth]{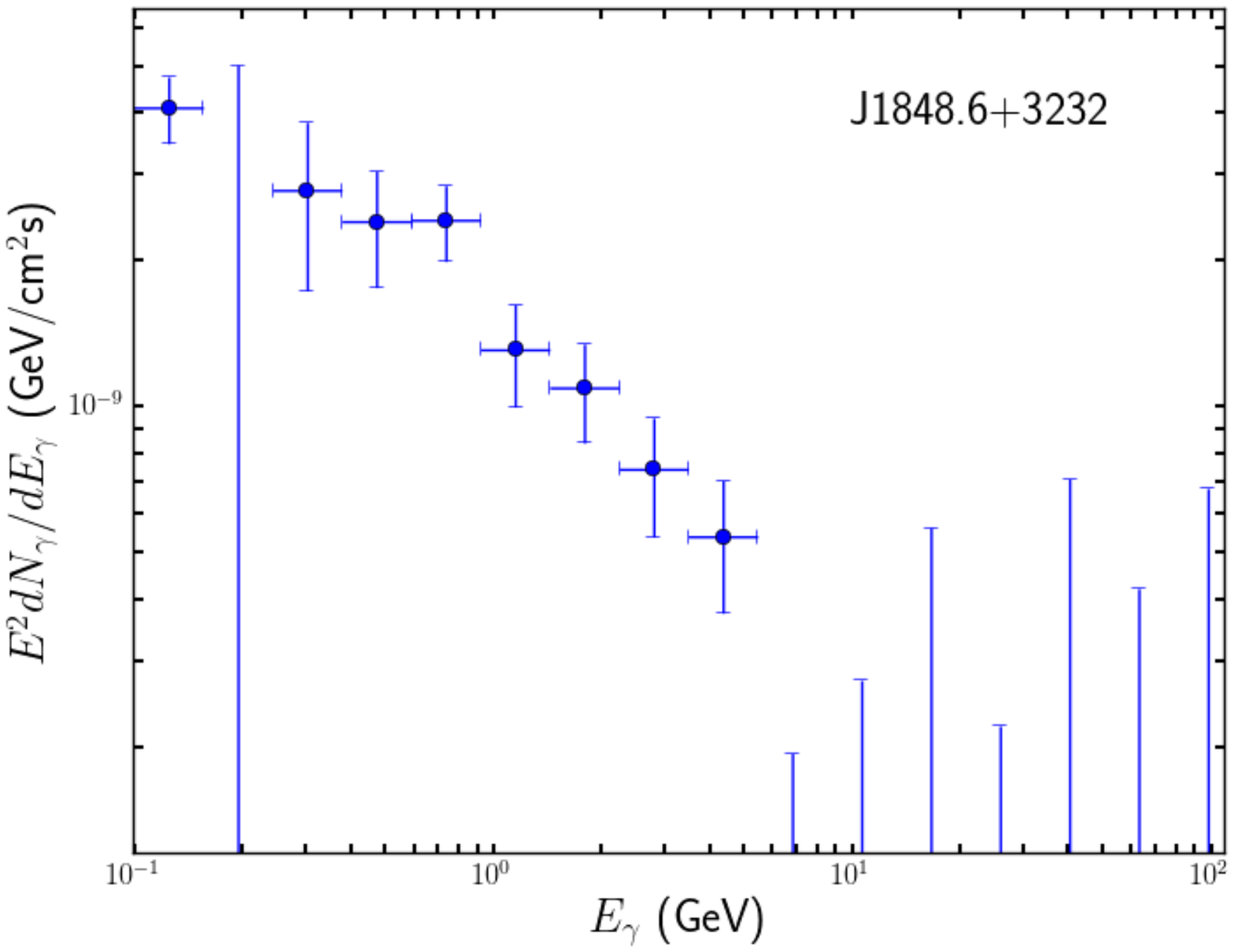}} \\
\vspace{-5.2cm}
\mbox{\includegraphics[width=0.49\textwidth]{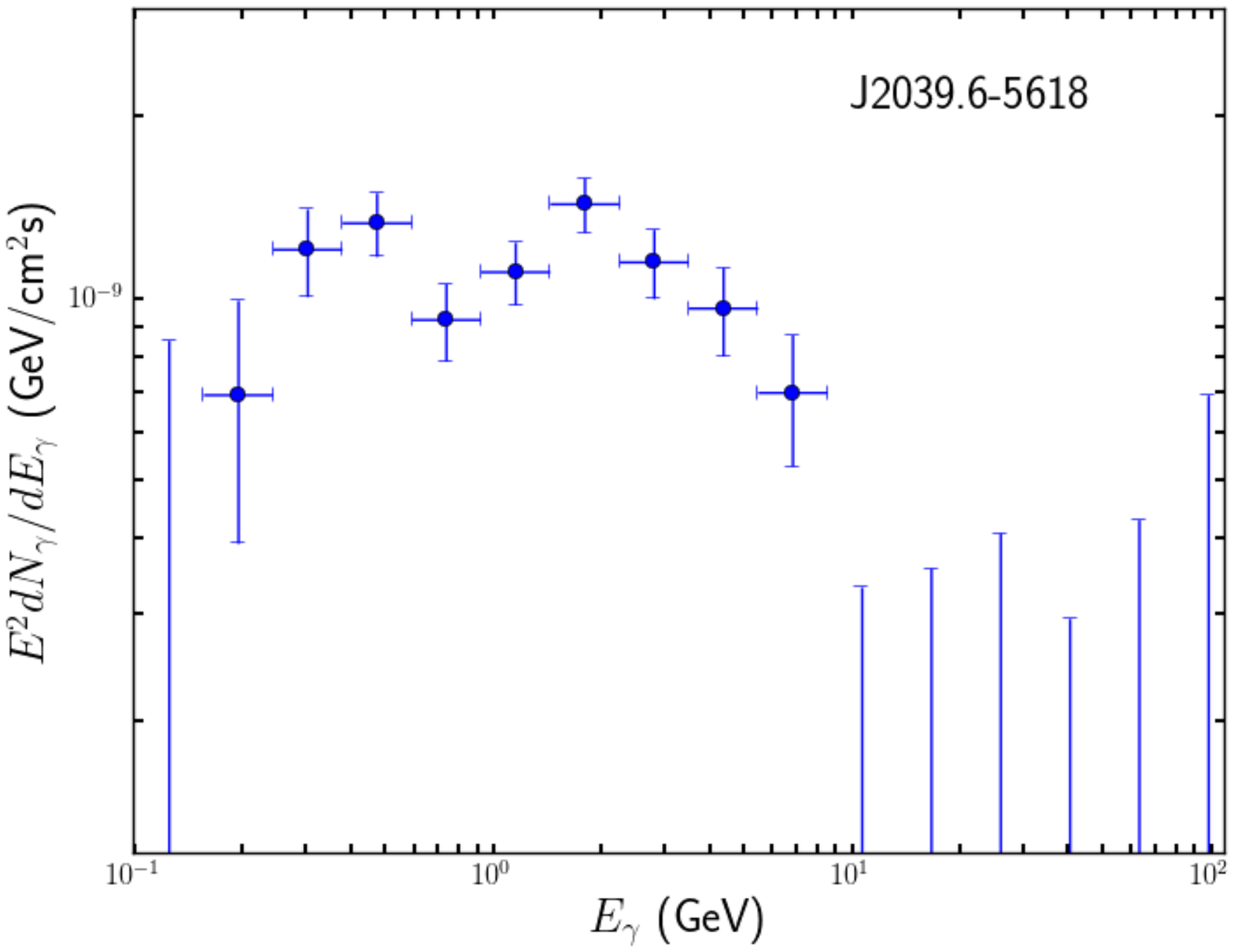}} \\
\vspace{-2.6cm}
\caption{As in Fig.~\ref{goodfits}, but for the 5 bright ($F_{\gamma} > 2 \times 10^{-9}$ cm$^{-2}$ s$^{-1}$, $E_{\gamma} > 1$ GeV) non-variable (variability index $< 80$), unassociated 3FGL sources located outside of the Galactic Plane ($|b|>5^{\circ}$) that are {\it not} well-fit by any dark matter particle annihilating to $b\bar{b}$.}
\label{notgoodfits}
\end{figure*}

\begin{figure}
\vspace{-3.0cm}
\mbox{\includegraphics[width=0.52\textwidth]{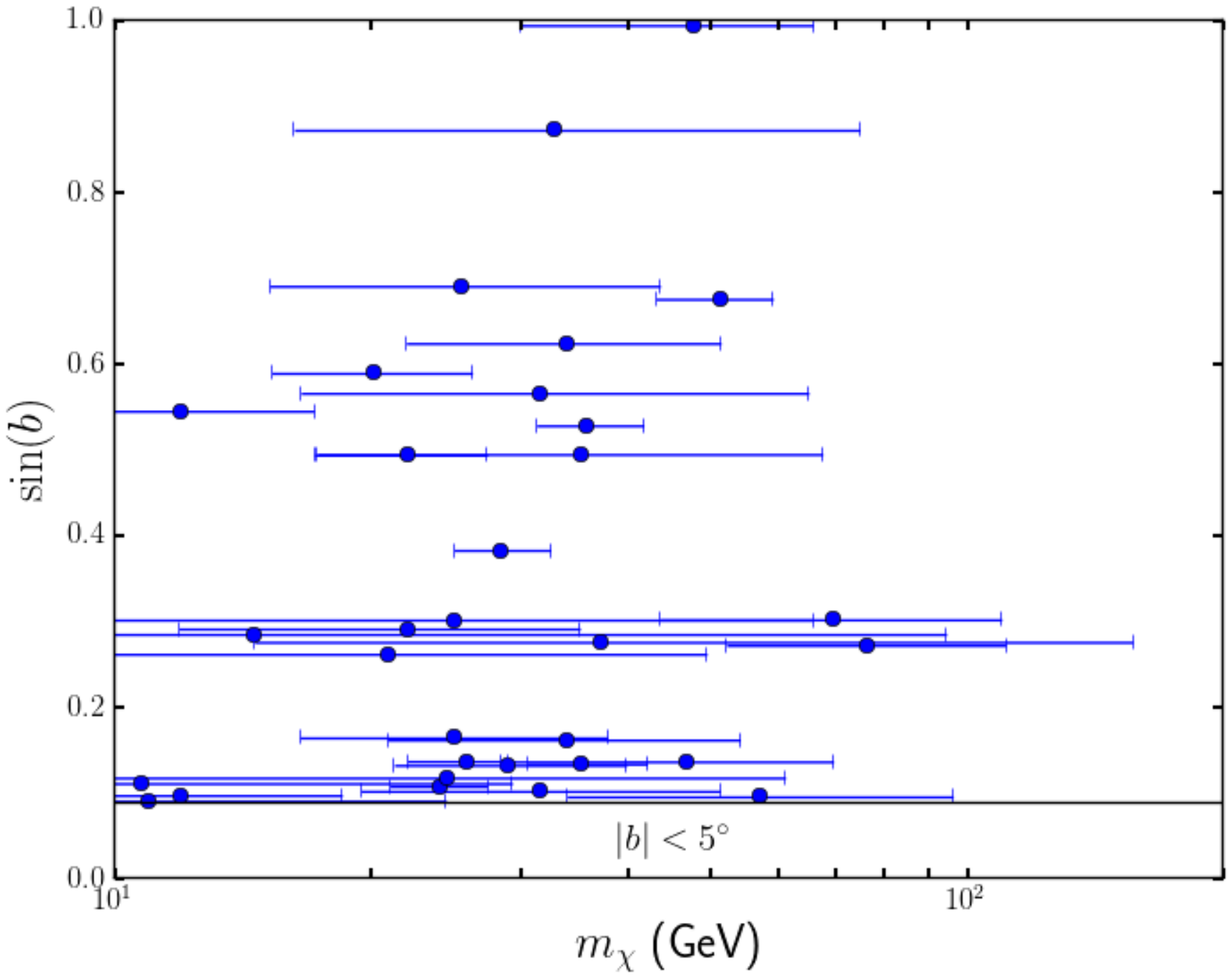}} 
\vspace{-3.0cm}
\caption{The distribution in galactic latitude and best-fit dark matter mass for those bright ($F_{\gamma} > 1.0 \times 10^{-9}$ cm$^{-2}$s$^{-1}$, $E_{\gamma} > 1$ GeV), non-variable (variability index $< 80$), unassociated 3FGL sources located outside of the Galactic Plane ($|b|>5^{\circ}$).  The error bars around each point represent the range of dark matter masses for which the fit to the spectrum yields $\chi^2 < 25$. In this figure, we have assumed annihilations that proceed to $b\bar{b}$.}
\label{isotropic}
\end{figure}

To calculate the gamma-ray spectra from Fermi's unassociated sources, we utilize data taken over approximately 6.4 years of Fermi-LAT observations,\footnote{MET range 239557417 - 442851954} utilizing the Pass 7 Reprocessed photons in the energy range of 100 MeV to 100 GeV. We exclude events arriving at a zenith angle greater than 100$^\circ$, as well as those which do not pass the ``Source'' photon data selection. We exclude events recorded while the instrument was not in science survey mode, when the instrumental rocking angle was $>$52$^\circ$, or when the instrument was passing through the South Atlantic Anomaly. We examine the photons observed within a 14$^\circ\times 14^\circ$ box centered around the location of each source, and divide the photons into 140$\times$140 angular bins and 15 evenly spaced logarithmic energy bins. We analyze the instrumental exposure throughout the region-of-interest using the {\tt P7REP\_SOURCE\_V15} instrumental response functions.

In our analysis, we employ the latest model for diffuse galactic gamma-ray emission, {\tt gll\_iem\_v05\_rev1.fit}, the latest isotropic emission template for the Source photon data selection {\tt iso\_source\_v05.txt}, and include all 3FGL sources which lie within our region-of-interest. We allow the normalization of each source to float independently in each energy bin, and do not impose any parameterization on their spectral shape. Our model also includes 3FGL sources which lie nearby (but outside) of the region-of-interest. To calculate the best fit flux from each source (in each energy bin), we use the Fermi-LAT {\tt gtlike} code, utilizing the {\tt MINUIT} algorithm. We also calculate the upper limit of the source flux using the {\tt pyLikelihood UpperLimits} tool in each energy bin, and present only the (2$\sigma$) upper limit whenever the flux is smaller than twice the calculated flux error. 

In Figs.~\ref{goodfits} and~\ref{notgoodfits}, we plot the gamma-ray spectra from the 14 brightest ($F_{\gamma} > 2 \times 10^{-9}$ cm$^{-2}$ s$^{-1}$, $E_{\gamma} > 1$ GeV) non-variable (variability index $< 80$), unassociated 3FGL sources located outside of the Galactic Plane ($|b|>5^{\circ}$). The 9 sources shown in Fig.~\ref{goodfits} are reasonably well fit ($\chi^2 < 25$) by dark matter annihilating to $b\bar{b}$, and the dashed curves represent this prediction for the best-fit value of the dark matter mass. The 5 sources shown in Fig.~\ref{notgoodfits}, in contrast, are not well-fit for any choice of the dark matter mass.

\begin{table*}[t]
\centering
\begin{tabular}{|c|c|c|c|c|c|}
	\hline
Source Name &  $\Phi_{\gamma}$ (cm$^{-2}$ s$^{-1}$)  & $b$ (deg)  & $m_{\chi}$ (GeV), best-fit &$m_{\chi}$ (GeV), $\Delta \chi^2=4$\\
\hline
 3FGL J0312.1-0921  &  9.49$\times 10^{-10}$ & -52.36    &    31.1   &     12.0 --        72.3   \\     
 3FGL J0318.1+0252 &   1.23$\times 10^{-9}$ & -43.64    &    25.5     &  19.5 --        32.2    \\   
  3FGL J0456.2-6924   & 7.62$\times 10^{-10}$  &-35.28     &   10.9     &   $<$        28.9   \\       
 3FGL J0953.7-1510  &  1.25$\times 10^{-9}$ &  29.61  &      35.2    &    26.9 --        47.8    \\     
 3FGL J1119.9-2204  &  2.70$\times 10^{-9}$ &  36.06    &    20.2   &     17.2 --        23.7       \\    
 3FGL J1120.6+0713 &   1.10$\times 10^{-9}$ &   60.69    &    32.8  &     23.3 --       50.5   \\      
3FGL J1221.5-0632  &  8.06$\times 10^{-10}$  & 55.55  &      28.9     &   17.8 --       47.8    \\  
 3FGL J1225.9+2953  &  1.42$\times 10^{-9}$ &  83.76  &      47.8   &     30.0 --       64.9  \\           
 3FGL J1315.7-0732  &  8.35$\times 10^{-10}$ &  54.83   &    34.6    &    17.8 --     77.7    \\       
  3FGL J1543.5-0244  &  7.35$\times 10^{-10}$  & 38.90  &      10.0      & $<$       40.7    \\     
 3FGL J1544.6-1125   & 1.01$\times 10^{-9}$ &  32.98     &   12.0  &      $<$    16.8      \\     
 3FGL J1601.9+2306 &   8.56$\times 10^{-10}$ &  46.94&      27.4   &     $<$        187.7     \\   
 3FGL J1625.1-0021  &  3.6$\times 10^{-9}$  & 31.84    &    35.9     &   31.6 -- 41.4    \\  
 3FGL J1659.0-0142  &  9.26$\times 10^{-10}$ &  23.91   &     37.9  &     $<$     273.8  \\      
 3FGL J1704.4-0528  &  8.89$\times 10^{-10}$ &  20.80   &     35.2    &   $<$       964.7     \\   
 3FGL J1720.7+0711  &  8.89$\times 10^{-10}$ &  23.41   &     22.9     &   $<$        72.3      \\     
 3FGL J1744.1-7619 &   3.85$\times 10^{-9}$ & -22.47   &     28.4  &      25.5 --       32.2     \\      
  3FGL J1803.3-6706  &  7.12$\times 10^{-10}$ & -20.37   &     45.3   &     25.0 --        72.4         \\  
  3FGL J1946.4-5403  &  1.72$\times 10^{-9}$ & -29.56   &     22.1   &    17.2 --        26.9        \\
 3FGL J2112.5-3044  &  3.26$\times 10^{-9}$ & -42.45      &  51.4    &    43.7 --        58.3    \\       
 3FGL J2134.5-2131 &   7.06$\times 10^{-10}$ & -45.08 &      50.5  &     17.5 --       165.5      \\   
 3FGL J2212.5+0703 &   1.24$\times 10^{-9}$&  -38.56  &      34.0     &   21.8 --       51.5       \\   
 3FGL J2103.7-1113  &  1.091$\times 10^{-9}$ & -34.42  &      31.6    &   21.3 --        47.0     \\       
  3FGL J2133.0-6433  &  8.36$\times 10^{-10}$ & -41.27   &     26.4   &     13.1 --        61.5      \\   
 		 \hline
	\end{tabular}
\caption{A list of Fermi's most promising dark matter subhalo candidates. In particular, this table includes all unassociated, non-variable 3FGL sources with $|b| > 20^{\circ}$, $\Phi_{\gamma} >7\times 10^{-10}$ cm$^{-2}$ s$^{-1}$ ($>$ 1 GeV), and that are well-fit ($\chi^2 <25$) by annihilating dark matter for at least one value of the dark matter's mass (assuming annihilations to $b\bar{b}$).}
\label{tab2}
\end{table*}

Of the well-fit sources shown in Fig.~\ref{goodfits}, we note that all 9 prefer dark matter masses in the range of roughly 20 to 70 GeV. This is further explored in Fig.~\ref{isotropic}, where we plot the galactic latitude and best-fit dark matter mass (for annihilations that proceed to $b\bar{b}$) for each of the unassociated 3FGL sources with $F_{\gamma} > 1.0 \times 10^{-9}$ cm$^{-2}$s$^{-1}$ ($E_{\gamma} > 1$ GeV), variability index $< 80$, and $|b|>5^{\circ}$. The error bars around each point represent the range of dark matter masses for which the fit to the spectrum yields $\chi^2 < 25$. We note two things about this plot.  First, the distribution of these sources is not isotropic, and a component concentrated around the disk is clearly evident. Among the 12 of these sources with $|b|>20^{\circ}$, however, the distribution is consistent with isotropy. Second, nearly all of these sources favor dark matter masses in the range of approximately 20 to 70 GeV. While this could represent an indication of dark matter subhalos with $m_{\chi} \sim 30-50$ GeV, the spectral shape in question is similar to that observed from many gamma-ray pulsars.

In Table~\ref{tab2}, we list what we consider to be Fermi's most promising dark matter subhalo candidates. This includes all unassociated and non-variable sources with $|b| > 20^{\circ}$, $\Phi_{\gamma} >7\times 10^{-10}$ cm$^{-2}$ s$^{-1}$ ($>$ 1 GeV), and that are well-fit ($\chi^2 <25$) by annihilating dark matter for at least one value of the dark matter's mass (assuming annihilations to $b\bar{b}$). In our opinion, the sources contained in this list merit further investigation.  If associated emission can be detected at other wavelengths, these sources could be excluded as subhalo candidates, allowing us to derive stronger constraints on the dark matter annihilation cross section (see Sec.~\ref{constraints}).  Alternatively, the lack of counterparts at other wavelengths could strengthen the case that one or more of these candidates are, in fact, dark matter subhalos.

\section{Spatial Extension}
\label{extension}

\begin{figure*}
\vspace{1.0cm}
\mbox{\includegraphics[width=1.0\textwidth]{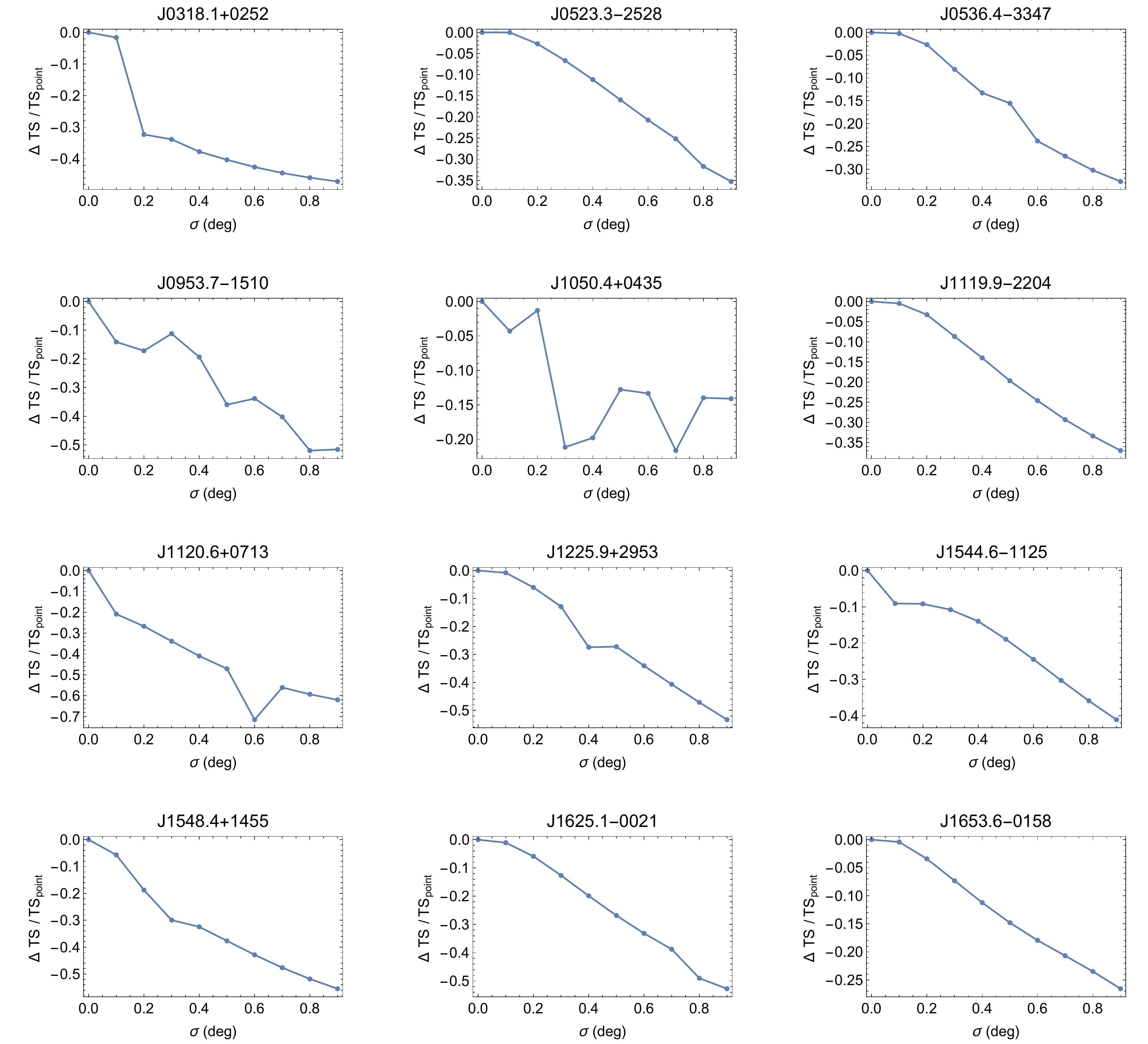}} 
\vspace{-0.3cm}
\caption{The change in the test statistic (TS) for Fermi's unassociated, very bright ($F_{\gamma} > 10^{-9}$ cm$^{-2}$ s$^{-1}$, $E_{\gamma} > 1$ GeV), high latitude ($|b|>20^{\circ}$) sources when the point-like template is replaced with a spatial template with gaussian extension.}
\label{ext}
\end{figure*}

\renewcommand{\thefigure}{\arabic{figure} (Cont.)}
\addtocounter{figure}{-1}

\begin{figure*}
\vspace{1.0cm}
\mbox{\includegraphics[width=1.0\textwidth]{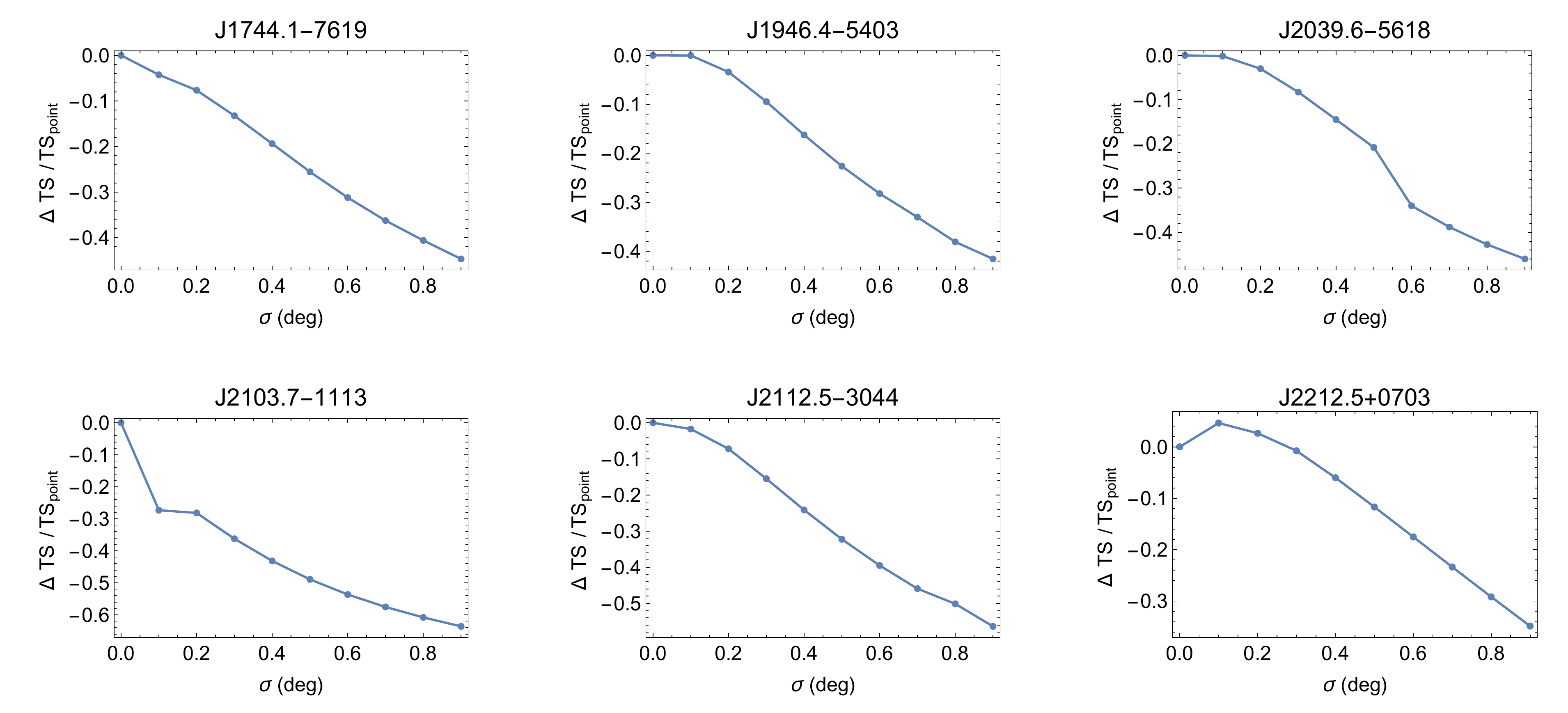}} 
\vspace{-0.3cm}
\caption{}
\end{figure*}

\renewcommand{\thefigure}{\arabic{figure}}

\begin{figure}
\mbox{\includegraphics[width=0.42\textwidth]{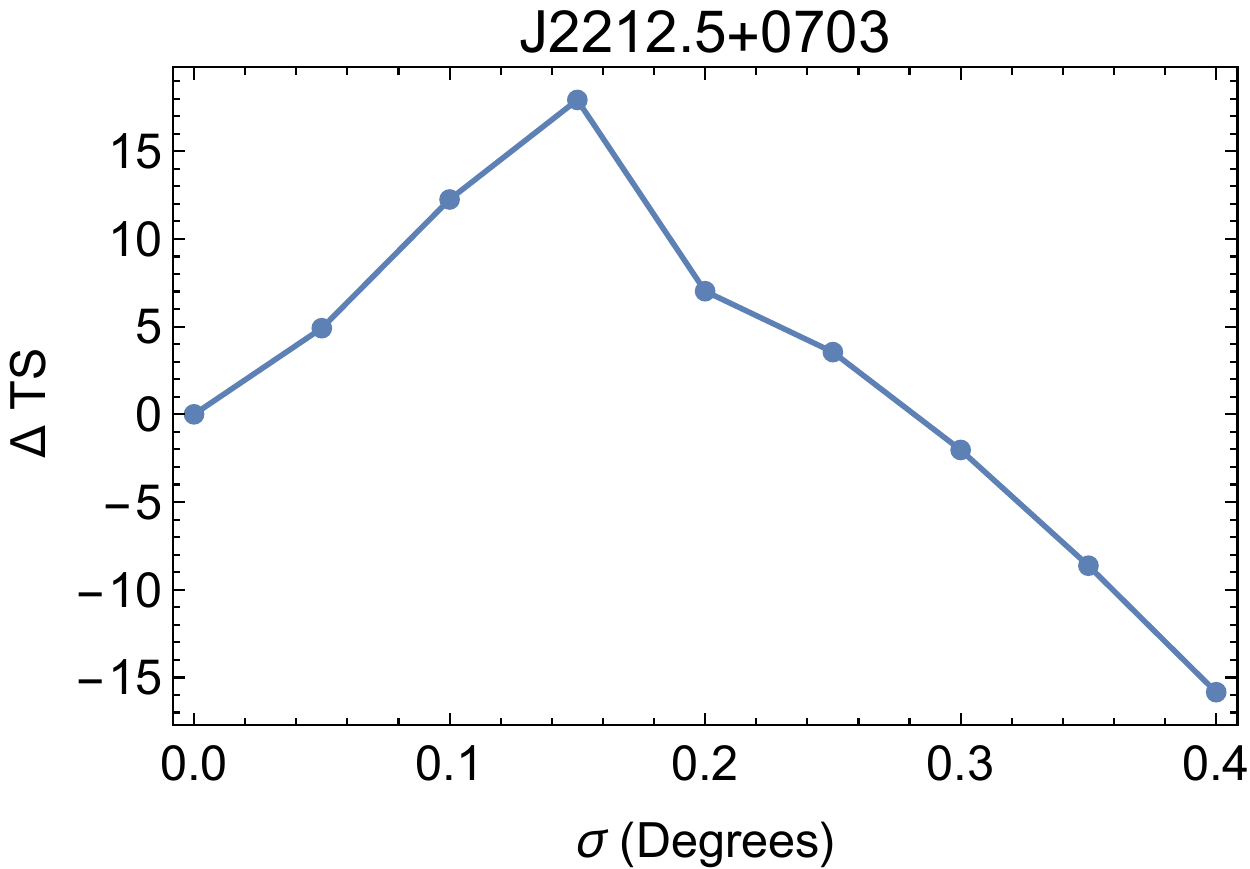}} \\
\caption{The change in the test statistic (TS) for Fermi's unassociated source J2212.5+0703 when the point-like template is replaced with a spatial template with gaussian extension.}
\label{J2212}
\end{figure}

\begin{figure}
\vspace{-3.0cm}
\mbox{\includegraphics[width=0.49\textwidth]{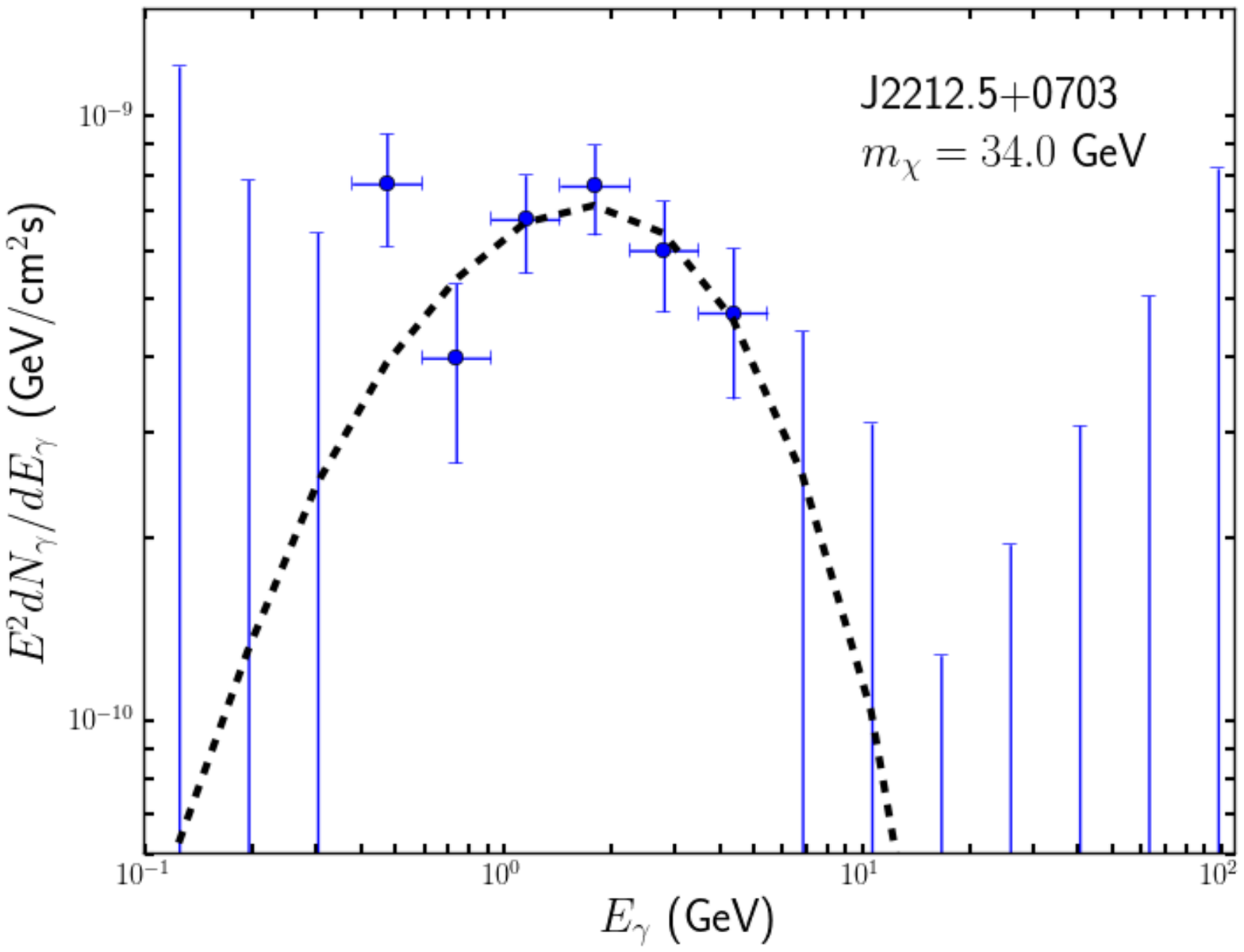}} 
\vspace{-3.0cm}
\caption{The spectrum of J2212.5+0703, compared to that predicted from a 34 GeV dark matter particle annihilating to $b\bar{b}$ (dashed curve).}
\label{J2212spec}
\end{figure}

\begin{figure*}
\vspace{1.0cm}
\mbox{\includegraphics[width=1.0\textwidth]{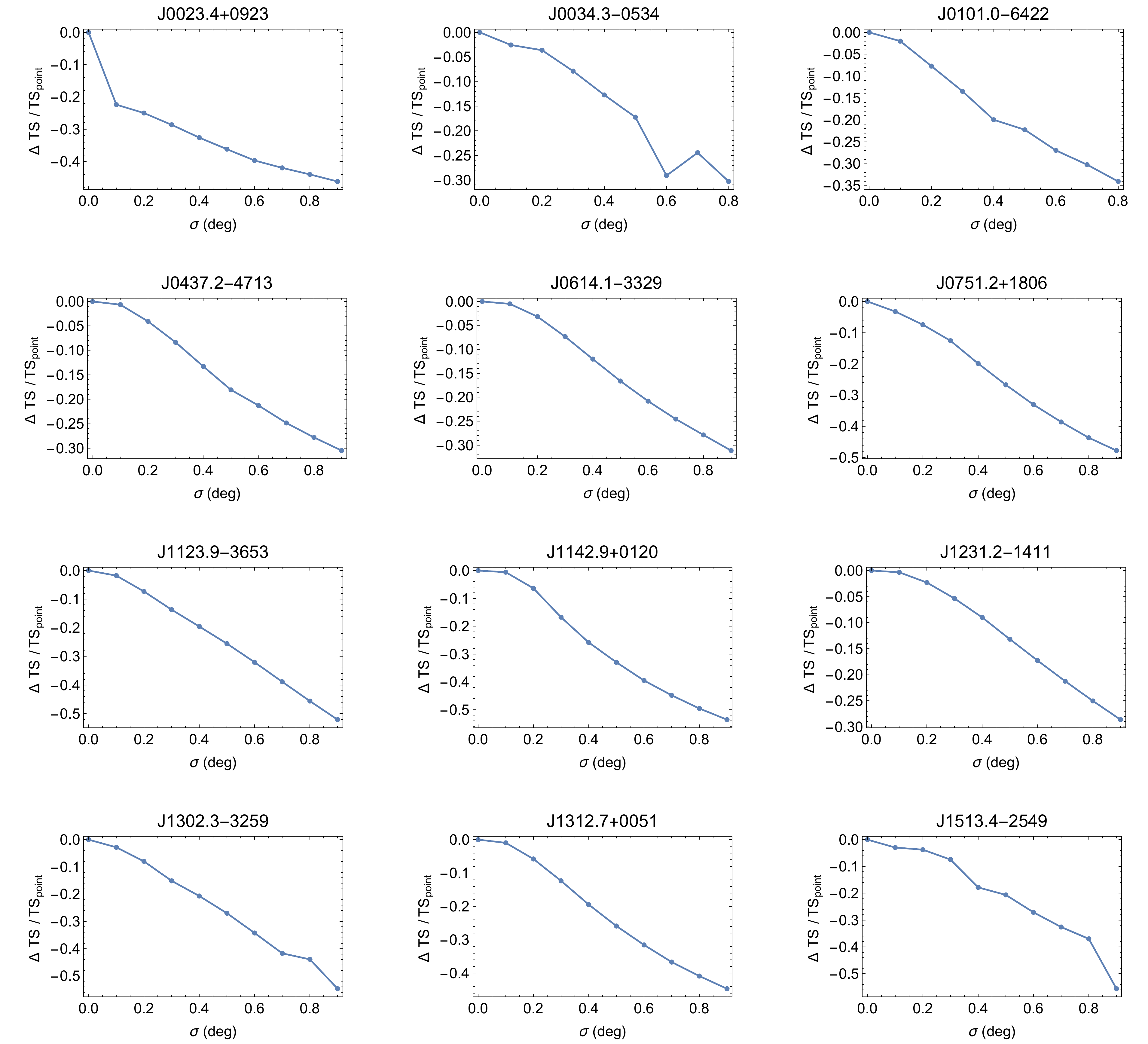}} 
\vspace{-0.3cm}
\caption{As in Fig.~\ref{ext}, but for Fermi sources that have been identified as pulsars that are located at high latitude ($|b|>20^{\circ}$) with a gamma-ray flux in the range of $F_{\gamma} = (1.0-4.3) \times 10^{-9}$ cm$^{-2}$ s$^{-1}$ ($E_{\gamma} > 1$ GeV). No evidence of spatial extension is found among this control group of sources.}
\label{extpulsar}
\end{figure*}

\renewcommand{\thefigure}{\arabic{figure} (Cont.)}
\addtocounter{figure}{-1}

\begin{figure*}
\vspace{1.0cm}
\mbox{\includegraphics[width=1.0\textwidth]{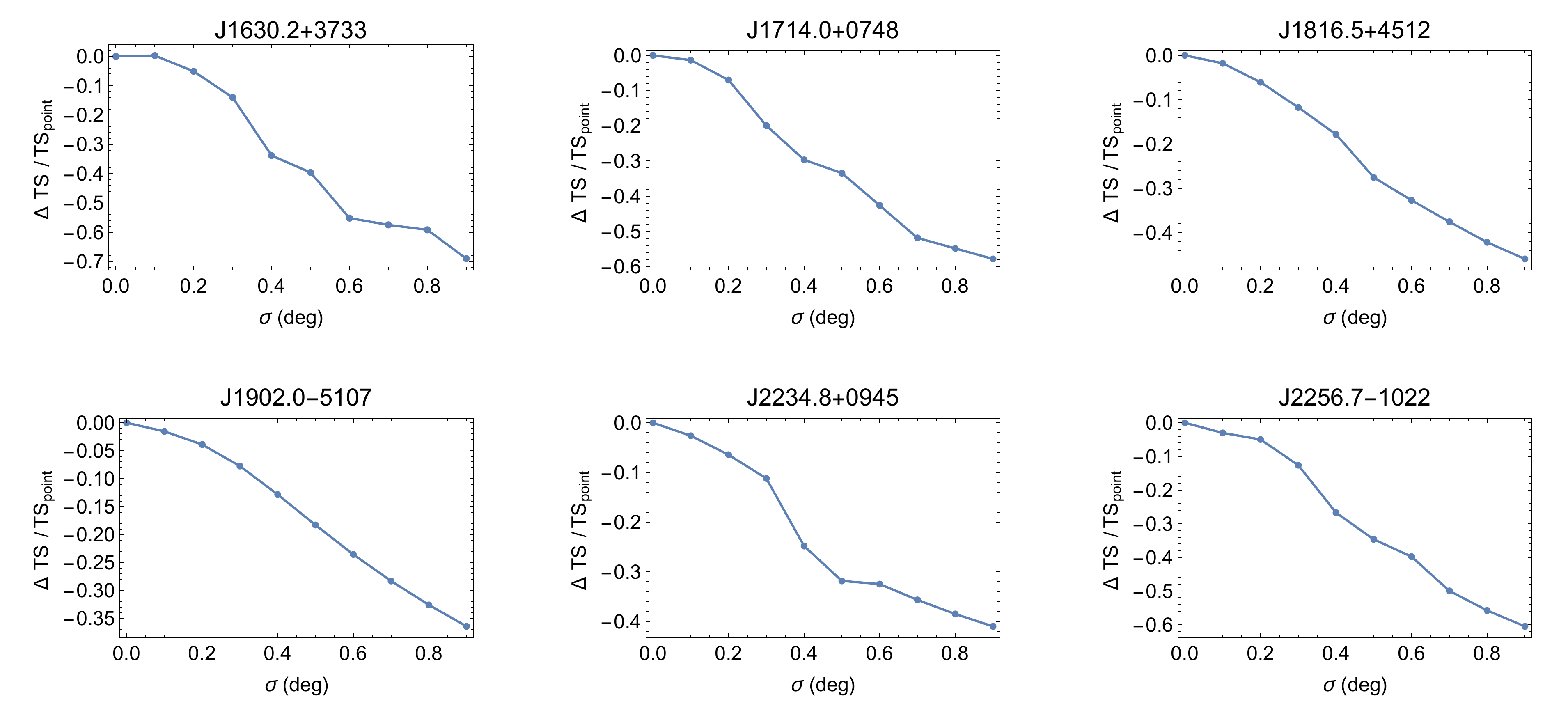}} 
\vspace{-0.3cm}
\caption{}
\end{figure*}

\renewcommand{\thefigure}{\arabic{figure}}

Most astrophysical gamma-ray emitters (pulsars, blazers, etc.) are effectively point sources, without any potentially observable spatial extension. A fraction of dark matter subhalos, on the other hand, could be detectably extended, especially those most nearby and large. In this section, we study the dark matter subhalo candidate sources identified in the previous section in an effort to determine whether they exhibit any evidence of spatial extension. 

The Fermi Collaboration has previously performed a search for spatial extension among their brightest sources, using their first two years of data~\cite{Lande:2012xn}. In doing so, they identified 21 spatially extended gamma-ray sources, none of which appear likely to be dark matter subhalos (17 of these 21 objects lie within 10 degrees of the Galactic Plane, and the four others are each associated with emission from the Large and Small Magellanic Clouds, the nearby galaxy Centaurus A, and the Ophiuchus molecular cloud). Given the fact that they did not identify other spatially extended sources, we do not expect any of the subhalo candidates under consideration to exhibit extension at a highly statistically significant level.  Given the larger data set that is currently available, however, it is possible that spatial extension could be detectable in one or more of our subhalo candidates at a non-negligible level.

To test for evidence of spatial extension, we repeat the procedure described in the previous section, replacing the point-source template with a gaussian template of width $\sigma$. In Fig.~\ref{ext}, we plot the change in the test statistic (TS) (defined as twice the difference in the global log-likelihood) of the source when the point-like template is replaced with that of an extended source.  Of these 18 sources considered (those with $|b|>20^{\circ}$ and $F_{\gamma} > 10^{-9}$ cm$^{-2}$ s$^{-1}$, $E_{\gamma} > 1$ GeV), we found one for which the TS increased when a spatially extended template was adopted (see Fig.~\ref{ext}).  This source, J2212.5+0703, is one of the promising high-latitude dark matter subhalo candidates identified in Sec.~\ref{data}, and is well fit by $m_{\chi} \simeq 21.8-51.5$ GeV (for annihilations to $b\bar{b}$); see Table~\ref{tab2}. We find that the TS of this source increases by 17.9 when the point-like template is replaced by a gaussian template of width $0.15^{\circ}$, corresponding to a local statistical significance of 4.2$\sigma$ (or 3.6$\sigma$ after accounting for a trials factor of 12, corresponding to the number of subhalo candidate sources tested for spatial extension). This is shown in greater detail in Fig.~\ref{J2212}. In Fig.~\ref{J2212spec}, we show the gamma-ray spectrum of this source. If the significance of this extension were to increase as Fermi continues to collect more data, it would help to support a dark matter subhalo interpretation for this source, over that of a pulsar or other point-like object. 

As a control group, we looked for signs of spatial extension among a subset of Fermi sources that have been identified as pulsars (see Fig.~\ref{extpulsar}).  In particular, we consider pulsars with the same range of latitudes ($|b|>20^{\circ}$) and gamma-ray fluxes ($F_{\gamma} = 1.0-4.3 \times 10^{-9}$ cm$^{-2}$ s$^{-1}$, $E_{\gamma} > 1$ GeV) as those subhalo candidates shown in Fig.~\ref{ext}. We find no evidence of spatial extension among this control group of sources (the most significant preference for extension was from J1630.2+3733, for which the TS increased by 0.6 when modeled with an extension of $\sigma=0.1^{\circ}$).
\\
\vspace{300.0cm}

\section{Constraining the Dark Matter Annihilation Cross Section}
\label{constraints}

\begin{figure*}
\vspace{-2.0cm}
\mbox{\includegraphics[width=0.52\textwidth]{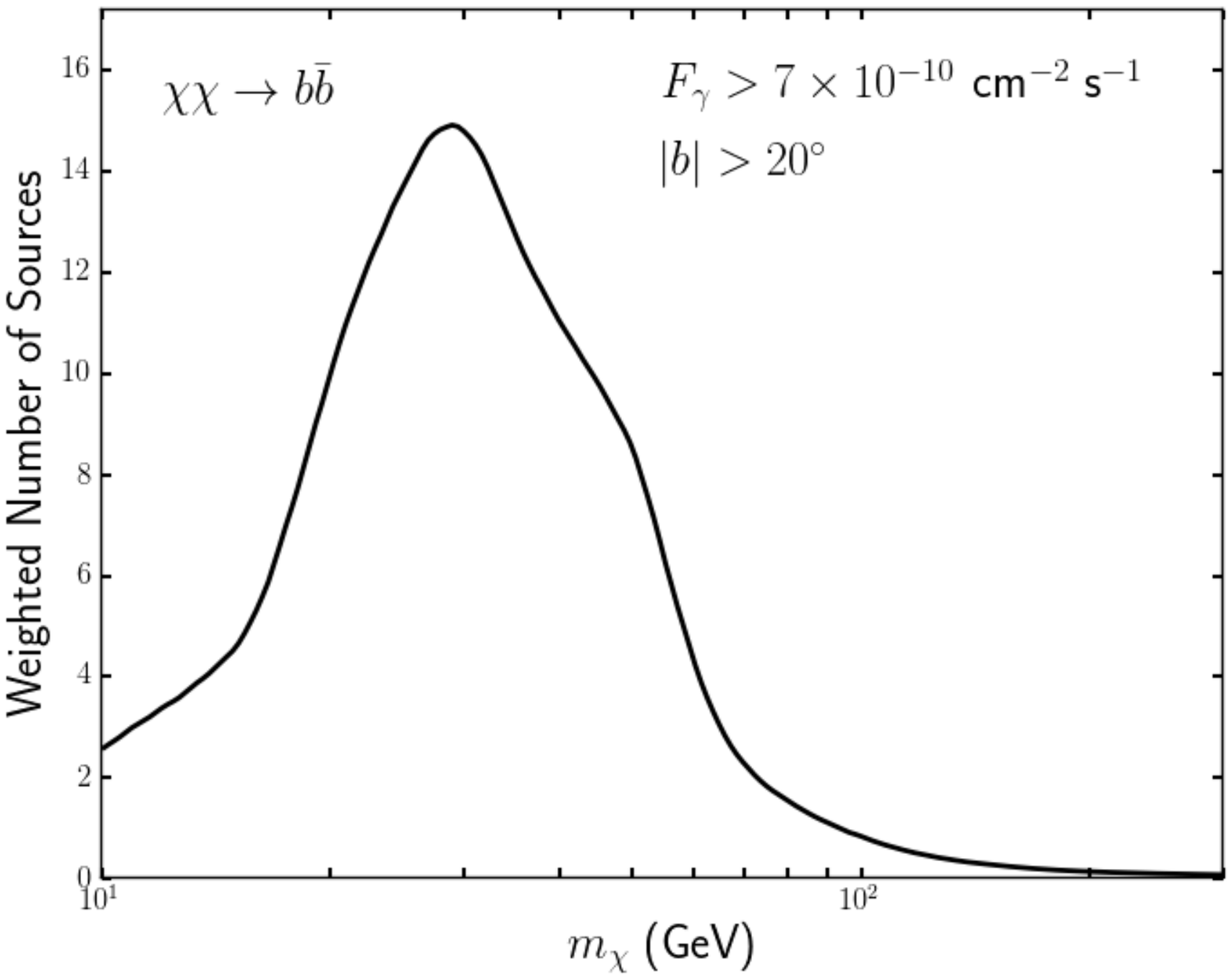}} 
\hspace{-1.0cm}
\mbox{\includegraphics[width=0.52\textwidth]{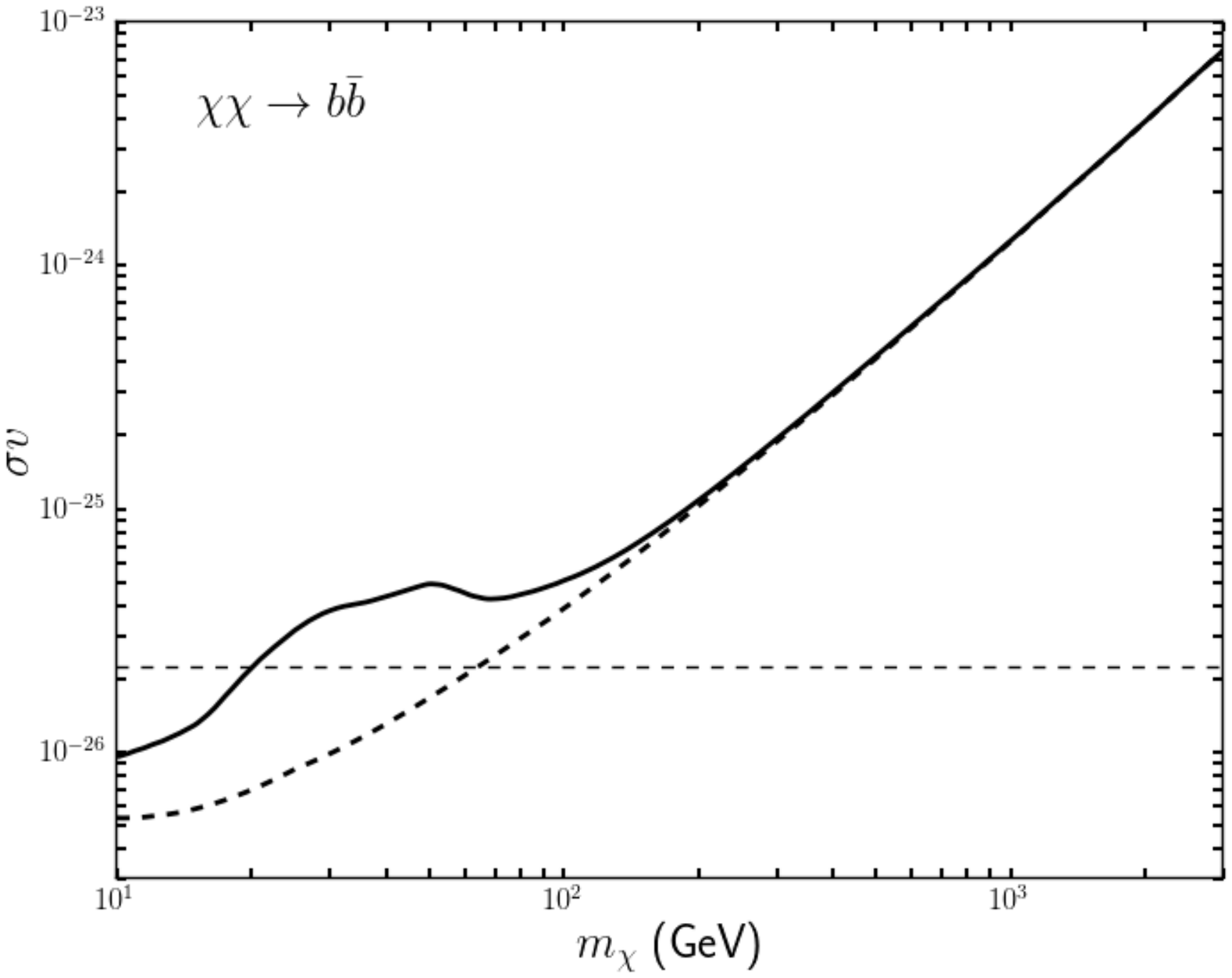}} 
\vspace{-2.6cm}
\caption{Left: The weighted number of unassociated, non-variable (variability index $<$ 80), bright ($F > 7 \times 10^{-10}$ cm$^{-2}$s$^{-1}$, $E_{\gamma}>1$ GeV), high-latitude ($|b| > 20^{\circ}$) 3FGL sources as a function of dark matter mass (assuming annihilations to $b\bar{b}$), defined as twice the sum of the $p$-values for all of the candidate sources under consideration. Right: The 95\% upper limit on the dark matter annihilation cross section derived from this source population. The dashed line represents the constraint that would have been derived if zero subhalo candidates had been observed. For high values of the dark matter mass, the weighted number of sources is only slightly greater than zero, and these two lines are almost identical.  For lower masses, in contrast, many subhalo candidates exist and the resulting constraint is somewhat weaker.}
\label{constraintfig1}
\end{figure*}

In this section, we use the results of Sec.~\ref{data} to place an upper limit on the number of bright, high-latitude dark matter subhalos, and use this to derive a constraint on the dark matter annihilation cross section. We do this through the following procedure. First, for each subhalo candidate, we calculate the $\chi^2$ of the fit for gamma-ray spectrum predicted by a given dark matter model, and convert this into a $p$-value. Twice the sum of the $p$-values for all of the candidate sources under consideration represents the weighted number of sources, which we plot in the left frame of Fig.~\ref{constraintfig1}. As already seen in Fig.~\ref{isotropic}, this result indicates that $\sim$10-15 of these sources could plausibly be subhalos of $\sim$20-50 GeV dark matter particles (they could also be pulsars, however).  Dark matter particles with heavier masses ($\gsim 100$ GeV), in contrast, do not provide a good-fit to any of the subhalo candidate sources. 

In the right frame of Fig.~\ref{constraintfig1}, we plot as a solid line the 95\% upper limit on the dark matter annihilation cross section derived from this source population. This result is obtained by calculating the Poisson errors around the weighted number of sources shown in the left frame. The dashed line represents the constraint that would have been derived if zero subhalo candidates had been observed. For high values of the dark matter mass, the weighted number of sources is only slightly greater than zero, and these two lines are almost identical.  For lower masses, in contrast, many subhalo candidates exist and the resulting constraint is somewhat weaker.

\begin{figure*}
\vspace{-2.0cm}
\mbox{\includegraphics[width=0.52\textwidth]{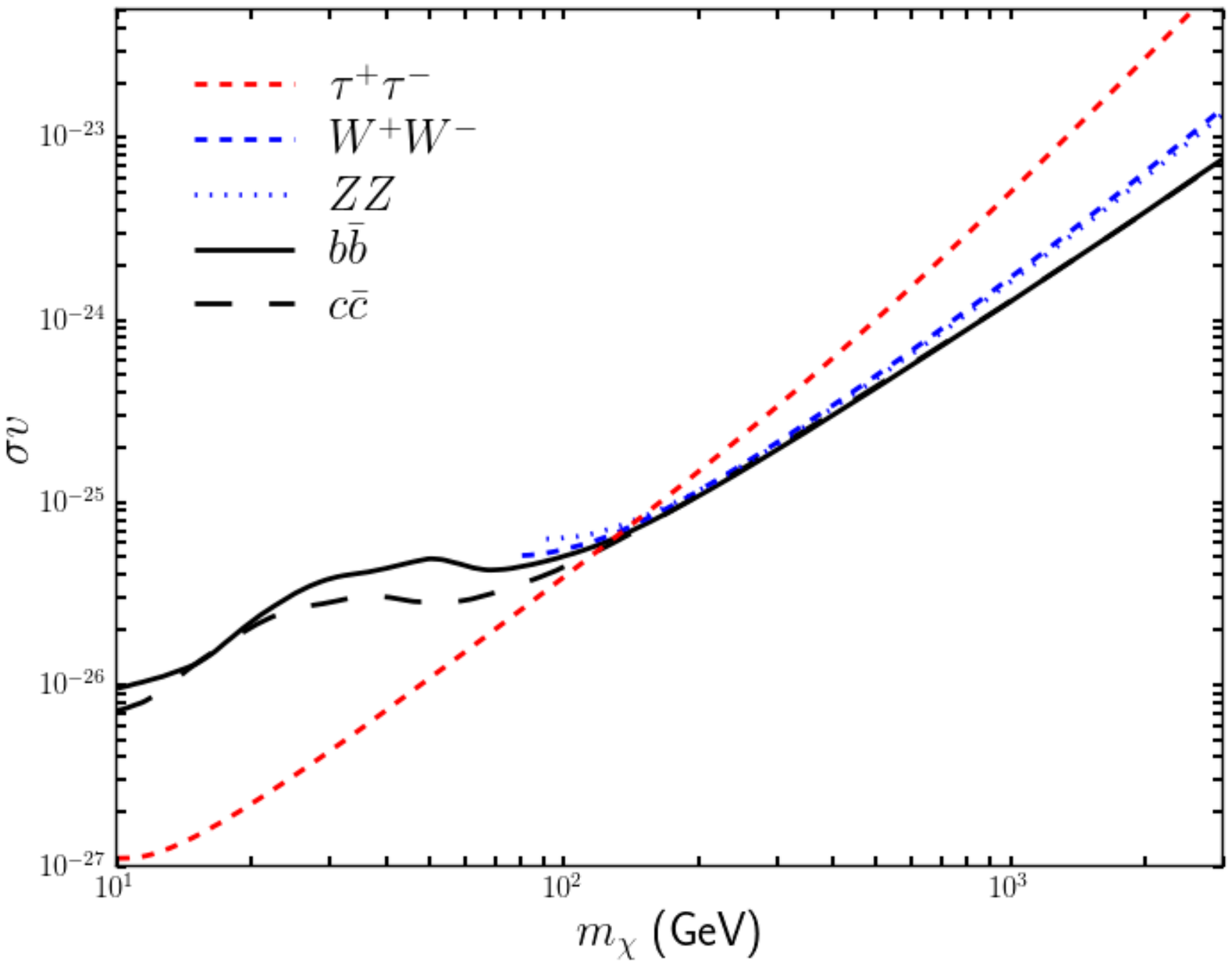}}
\hspace{-1.0cm}
\mbox{\includegraphics[width=0.52\textwidth]{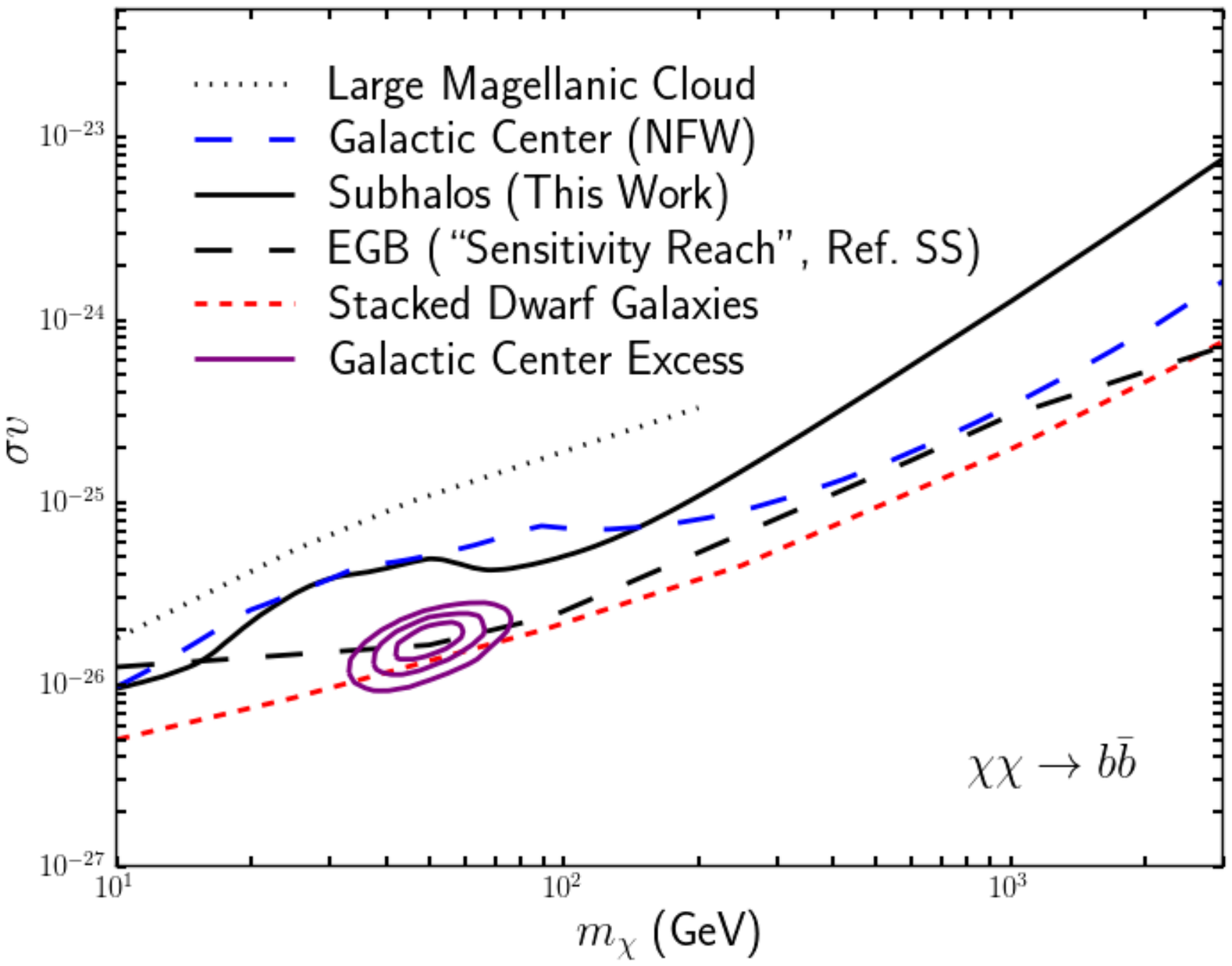}}
\vspace{-2.6cm}
\caption{Left: The 95\% upper limit on the dark matter annihilation cross section, for a variety of annihilation channels. Right: A comparison of the limit presented here to those derived from gamma-ray observations of dwarf spheroidal galaxies~\cite{Drlica-Wagner:2015xua}, the Galactic Center~\cite{Hooper:2012sr}, the extragalactic gamma-ray background~\cite{Ackermann:2015tah}, and the Large Magellanic Cloud~\cite{Buckley:2015doa}. Also shown is the region of parameter space that is able to account for the gamma-ray excess observed from the region surrounding the Galactic Center~\cite{Goodenough:2009gk,Hooper:2010mq,Abazajian:2012pn,Gordon:2013vta,Hooper:2013rwa,Abazajian:2014fta,Daylan:2014rsa}, as presented in Refs.~\cite{Calore:2014xka,Calore:2014nla} (assuming a local dark matter density of 0.4 GeV/cm$^3$ and a scale radius for the Milky Way's halo profile of 20 kpc).}
\label{constraintfig2}
\end{figure*}

In the left frame of Fig.~\ref{constraintfig2}, we show this constraint as derived for a variety of dark matter annihilation channels ($\tau^+ \tau^-$, $W^+ W^-$, $ZZ$, $b\bar{b}$, $c\bar{c}$). We do not consider annihilations to $e^+e^-$, $\mu^+ \mu^-$, as these final states are more strongly constrained by cosmic ray positron measurements~\cite{Bergstrom:2013jra} than by gamma-ray observations. In the right frame of this figure, we compare the constraints presented in this paper to those derived from gamma-ray observations of dwarf spheroidal galaxies~\cite{Drlica-Wagner:2015xua} (see also, Ref.~\cite{Geringer-Sameth:2014qqa}), the Galactic Center (assuming an NFW halo profile)~\cite{Hooper:2012sr}, the extragalactic gamma-ray background (assuming a reference substructure model, and taking the less conservative ``sensitivity reach'' approach)~\cite{Ackermann:2015tah} (see also, Ref.~\cite{DiMauro:2015tfa}), and the Large Magellanic Cloud~\cite{Buckley:2015doa}. The constraints presented here are comparably stringent to those derived from these other observations, and are complementary due to the differing systematic uncertainties involved. 

We also show in the right frame of Fig.~\ref{constraintfig2} the region of parameter space that is able to account for the gamma-ray excess observed from the region surrounding the Galactic Center~\cite{Goodenough:2009gk,Hooper:2010mq,Abazajian:2012pn,Gordon:2013vta,Hooper:2013rwa,Abazajian:2014fta,Daylan:2014rsa}, as presented in Refs.~\cite{Calore:2014xka,Calore:2014nla}. This region assumes a local dark matter density of 0.4 GeV/cm$^3$ and a scale radius for the Milky Way's halo profile of 20 kpc. Varying these quantities within their current uncertainties can allow this region to shift upward or downward by an additional factor of a few. It is interesting to note that if the Galactic Center excess is generated by annihilating dark matter, then we expect that a few of our dark matter subhalo candidate sources should, in fact, be dark matter subhalos.

\section{Summary and Prospects}
\label{summary}

The dark matter halo of the Milky Way is predicted to contain a very large number of smaller subhalos, and if the dark matter consists of weak-scale annihilating particles, the most massive and nearby of these subhalos could appear as gamma-ray sources without counterparts at other wavelengths.  In this paper, we have studied the 992 unassociated gamma-ray sources contained in the Fermi Collaboration's Third Source Catalog (3FGL), in an attempt to constrain the number of these sources that could be dark matter subhalos. From this information, we have derived constraints on the dark matter annihilation cross section that are comparably stringent to those based on gamma-ray observations of dwarf spheroidal galaxies, the Galactic Center, and the extragalactic gamma-ray background. 

We also present in this paper a list of 24 sources which we consider to be promising dark matter subhalo candidates. These sources are each quite bright ($F_{\gamma} > 7\times 10^{-10}$ cm$^{-2}$ s$^{-1}$, $E_{\gamma} > 1$ GeV), are located far away from the Galactic Plane ($|b|>20^{\circ}$), show no signs of variability (variability index $<80$), and exhibit a spectral shape that is compatible with that predicted from annihilating dark matter particles. The majority of these subhalo candidate sources are best-fit by dark matter particles with masses in the range of $\sim$20-50 GeV (assuming annihilations to $b\bar{b}$), similar to that favored to explain the previously reported Galactic Center gamma-ray excess~\cite{Goodenough:2009gk,Hooper:2010mq,Abazajian:2012pn,Gordon:2013vta,Hooper:2013rwa,Abazajian:2014fta,Daylan:2014rsa,Calore:2014xka,Calore:2014nla}. Gamma-ray pulsars are known to exhibit similar spectra, however, making it difficult to determine whether these sources are dark matter subhalos or radio-faint gamma-ray pulsars. We also report 3.6$\sigma$ (global) evidence of spatial extension from the bright, high-latitude subhalo candidate source 3FGL J2212.5+0703.  If the significance of this extension continues to increase as Fermi collects more data, and no multi-wavelength counterparts are identified, it would strengthen the case that this source could be a dark matter subhalo, rather than a pulsar or other point-like gamma-ray source. 

Of the 117 high-confidence gamma-ray pulsars described in the Second Fermi-LAT Pulsar Catalog~\cite{TheFermi-LAT:2013ssa}, 36 have pulsations that were first identified in gamma-rays (as opposed to in radio or other wavelengths). Each of these sources has been studied with deep radio observations~\cite{Parkinson:2010xf,Ray:2010ws,Ray:2012ue}, yielding several detections~\cite{Camilo:2009ju,Abdo:2010ht,Pletsch:2011kp}.\footnote{For an updated list, see \url{https://confluence.slac.stanford.edu/display/GLAMCOG/Public+List+of+LAT-Detected+Gamma-Ray+Pulsars}} The fact that dozens of gamma-ray pulsars have not yet been detected at other wavelengths, however, leads us to expect that some of Fermi's unassociated sources could be radio-quiet pulsars, whose gamma-ray pulsations have thus far gone undetected (perhaps due to weak pulsation and/or broad pulse profiles). Deeper multi-wavelength observations and further searches for gamma-ray pulsations will be essential if we are to identify the nature of these intriguing sources.


\bigskip

{\it Acknowledgements}:  We would like to thank Alex Drlica-Wagner for valuable discussions. BB is supported by the US Department of Energy Office of Science Graduate Student Research (SCGSR) Program under Contrast No. DE-AC05-06OR23100. DH is supported by the US Department of Energy under contract DE-FG02-13ER41958. Fermilab is operated by Fermi Research Alliance, LLC, under Contract No. DE- AC02-07CH11359 with the US Department of Energy. TL is supported by the National Aeronautics and Space Administration through Einstein Postdoctoral Fellowship Award No. PF3-140110.

\bibliography{3fgl}

\end{document}